\documentclass[prb,aps, twocolumn, amsmath,amssymb,superscriptaddress]{revtex4}
\usepackage{float}
\usepackage{amsmath}
\usepackage{amssymb}
\usepackage{amsfonts}
\usepackage{euscript}
\usepackage{enumerate}
\usepackage{hhline}
\usepackage{pslatex}
\usepackage{tabularx}
\usepackage[usenames,dvipsnames]{xcolor}
\usepackage{graphicx}% Include figure files
\usepackage{dcolumn}% Align table columns on decimal point
\usepackage{bm}% bold math
\usepackage[sort&compress]{natbib}

\newcommand{\ket}[1] {| #1 \rangle}
\makeatletter
\renewcommand{\@biblabel}[1]{#1. }
\renewcommand{\@dotsep}{500}
\renewcommand{\@pnumwidth}{0em}
\renewcommand{\l@figure}[2]{% #1 is e.g. Figure 1 + caption, #2 is pg.
\@dottedtocline{1}{1.5em}{2em}{Figure #1}{}\vspace{15pt}}

\begin{document}

\title{Efficient and low-noise single-photon-level frequency conversion interfaces using silicon nanophotonics}

\author{Qing Li}\email{qing.li@nist.gov}
\affiliation{Center for Nanoscale Science and Technology, National
Institute of Standards and Technology, Gaithersburg, MD 20899,
USA}\affiliation{Maryland NanoCenter, University of Maryland,
College Park, MD 20742, USA}
\author{Marcelo Davan\c co}
\affiliation{Center for Nanoscale Science and Technology, National
Institute of Standards and Technology, Gaithersburg, MD 20899,
USA}
\author{Kartik Srinivasan} \email{kartik.srinivasan@nist.gov}
\affiliation{Center for Nanoscale Science and Technology, National
Institute of Standards and Technology, Gaithersburg, MD 20899, USA}

\date{\today}% It is always \today, today,
%  but any date may be explicitly specified

\begin{abstract}
\noindent \textbf{Optical frequency conversion has applications ranging from tunable light sources to telecommunications-band interfaces for quantum information science.  Here, we demonstrate efficient, low-noise frequency conversion on a nanophotonic chip through four-wave-mixing Bragg scattering in compact (footprint $<$~0.5~$\times$~10$^{-4}$~cm$^2$) Si$_3$N$_4$ microring resonators.  We investigate three frequency conversion configurations: (1) spectral translation over a few nanometers within the 980~nm band, (2) upconversion from 1550~nm to 980~nm, and (3) downconversion from 980~nm to 1550~nm.  With conversion efficiencies ranging from 25~$\%$ for the first process to $>60~\%$ for the last two processes, a signal conversion bandwidth $>$~1~GHz, $<60$ mW of continuous-wave pump power needed, and background noise levels between a few fW and a few pW, these devices are suitable for quantum frequency conversion of single photon states from InAs quantum dots. Simulations based on coupled mode equations and the Lugiato-Lefever equation are used to model device performance, and show quantitative agreement with measurements.}
\end{abstract}

 \maketitle

\maketitle

Efficient and low-noise frequency conversion has many applications in classical and quantum photonics, for example, casting signals to specific spectral channels for wavelength multiplexing, accessing spectral regions for which high-performance detectors are available, and connecting visible quantum memories with the low-loss telecommunications band. Years of progress in developing frequency conversion technology in both $\chi^{(2)}$ and $\chi^{(3)}$ nonlinear media~\cite{ref:Boyd_NLO_book,ref:Agrawal_NFO}, typically based on centimeter long crystals~\cite{ref:Langrock_Fejer} and meter long optical fibers~\cite{ref:Gnauck}, has enabled researchers to demonstrate frequency conversion of quantum states, or quantum frequency conversion (QFC)~\cite{ref:Kumar_OL}, down to the level of single photon Fock states~\cite{ref:Raymer_Srinivasan_PT_QFC}.

Nanophotonic geometries are an attractive way to miniaturize, scale, and customize optical components compared to conventional technology. In particular, microresonators with high quality factor ($Q$) and small mode volume have proven to be an excellent choice for nonlinear applications by offering small power budget and device size. Nonlinear wave mixing in the silicon platform (including silicon nitride and silicon dioxide) is dominated by the $\chi^{(3)}$ nonlinearity~\cite{ref:Moss_Gaeta_Lison_NLO}, which is also the dominant parametric nonlinearity in optical-fiber-based technology. Most previous $\chi^{(3)}$-based frequency conversion work in nanophotonic devices~\cite{ref:Foster_Gaeta,ref:Kippenberg_comb,ref:Turner_Lipson_FC_Si_rings,ref:Levy_Lipson_comb,Ref:Razzari_Hydex_comb,ref:Liu_Green_midIR_PA,ref:Zlatanovic_Radic_midIR_PA} has employed a degenerate or nondegenerate four-wave mixing (FWM) process called parametric amplification (PA), where idler generation goes along with vacuum noise amplification in a system with anomalous dispersion at the pump wavelength. While PA-based frequency conversion works well for many classical applications, it has theoretically been shown that there is a fundamental limit on the achievable signal-to-noise ratio (SNR) for the converted idler at a single-photon-level input, limiting its application for QFC~\cite{ref:Mckinstrie_parametric_processes}. Alternatively, there is a non-degenerate FWM process, termed four-wave-mixing Bragg scattering (FWM-BS)~\cite{ref:Marhic_Kazovsky_WE,ref:Uesaka_Kazovksy,ref:McKinstrie_FWM_BS}, that is typically operated in the normal dispersion regime and does not amplify vacuum fluctuations. As a result, FWM-BS can in principle be a noiseless process, and in fact its Hamiltonian has the form of a state transfer process and can be considered as an active beam splitter~\cite{ref:McKinstrie_FWM_BS}. In addition, PA does not allow an arbitrary choice of the target conversion wavelength, which can be important for both the aforementioned classical and quantum applications.  In contrast, FWM-BS provides a controllable spectral translation set by the difference in pump frequencies~\cite{ref:Marhic_Kazovsky_WE,ref:Eggleton_FWBS_theory_PRA}. While two earlier works demonstrated FWM-BS in nanophotonic waveguides, the efficiency was $\lesssim5~\%$, despite the use of short pulsed pumps with a peak power of a few Watts~\cite{ref:Agha_OL_FWM_BS,ref:Agha_OE_FWM_BS}. Recent theoretical works ~\cite{ref:Kumar_FWBS_theory_OL,ref:Loncar_FWBS_theory_OE} indicate a rising interest in developing FWM-BS using a microresonator structure, in order to achieve efficient single-photon-level frequency conversion with a reasonable power budget.

\begin{center}
\begin{figure}
\begin{center}
\includegraphics[width=\linewidth]{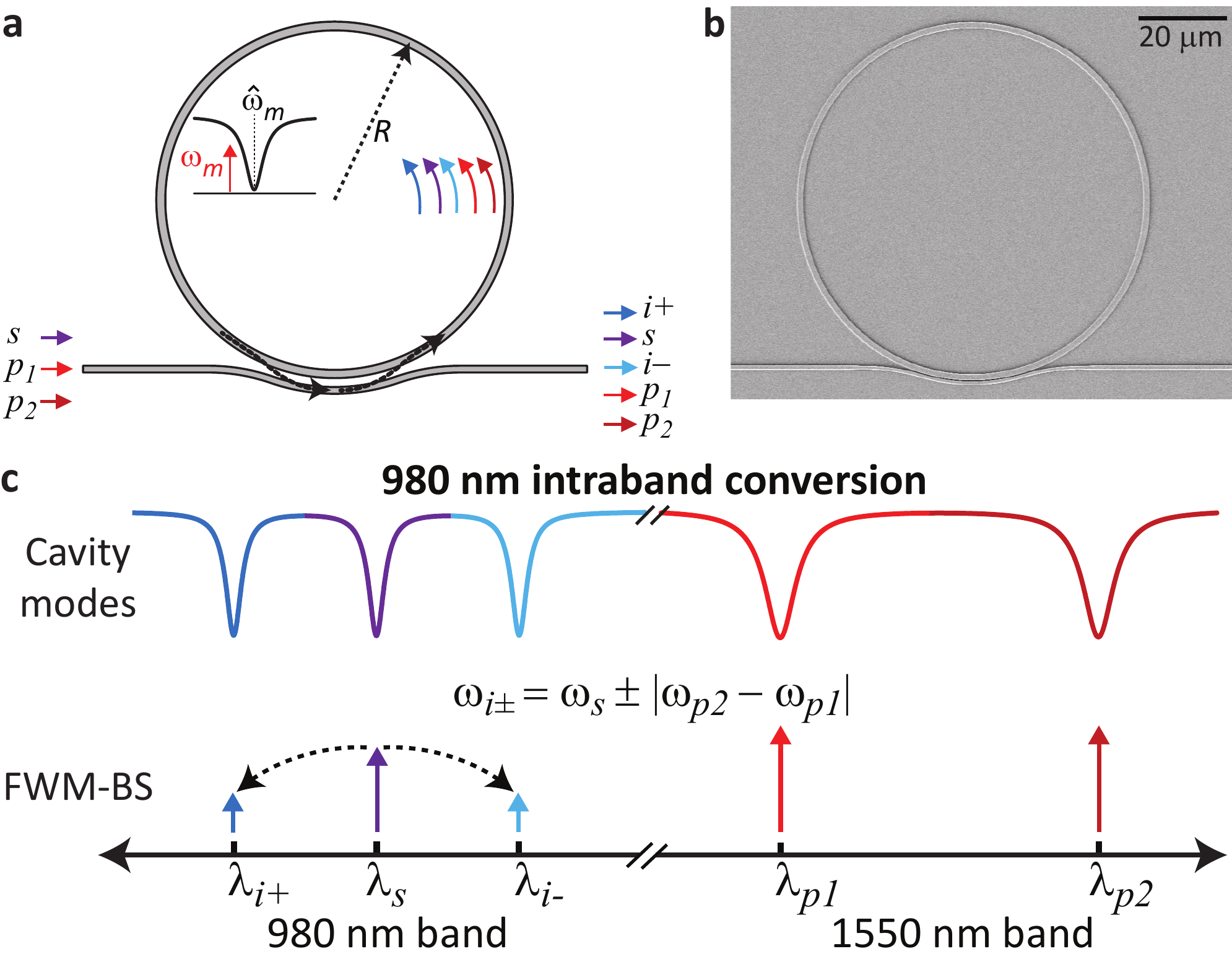}
\caption{\textbf{Principle of operation and device geometry}. \textbf{a}, Schematic of the microresonator geometry used, where an input signal $s$ is converted to idlers $i+$ and $i-$ through application of two pump fields $p_{1}$ and $p_{2}$. \textbf{b}, Scanning electron micrograph of a $40$ $\mu$m radius microring resonator, showing the pulley-coupled access waveguide. \textbf{c}, Schematic of the FWM-BS process for $980$ nm intraband frequency conversion, where $s$, $i+$, and $i-$ are in the 980~nm band and the pumps $p_{1}$ and $p_{2}$ are in the 1550 nm band.  Of critical importance to efficient conversion is the spectral matching of the frequency components generated by FWM-BS ($\omega_{s,i+,i-,p1,p2}$) with the resonant frequencies of the cavity modes ($\hat{\omega}_{s,i+,i-,p1,p2}$). As shown in Figs.~5-6, the same microring can be used to create a frequency conversion interface between the 980~nm and 1550~nm bands.}
\label{fig:Fig1}
\end{center}
\end{figure}
\end{center}

In this Article, we show for the first time a resonance-enhanced FWM-BS process, using a compact $\text{Si}_3\text{N}_4$ microring resonator that significantly reduces the required pump power ($< 60$ mW continuous-wave pumps). The $1550$ nm band and the $980$ nm band are considered due to their relevance to low-loss transmission through optical fibers and quantum light generation by InAs/GaAs quantum dots, respectively. We first show efficient ($>25~\%$ conversion efficiency) 980~nm intraband frequency conversion with spectral translation over a few nanometers, which can be used to achieve indistinguishability between InAs/GaAs quantum dots emitting with slightly different wavelengths. Then, we demonstrate wideband frequency conversion from the $1550$ nm band to the $980$ nm band (upconversion) and the $980$ nm band to the $1550$ nm band (downconversion), with each showing $>60~\%$ conversion efficiency. The background noise properties for both $980$ nm intraband conversion and wideband 980~nm to 1550~nm downconversion are studied using single-photon-level inputs. The performance level of our nanophotonic frequency conversion interfaces (in terms of conversion efficiency and SNR) is comparable to more mature $\chi^{(3)}$ technology such as highly nonlinear and photonic crystal fibers~\cite{ref:Gnauck,ref:McGuinnes_PRL10,ref:Clark_high_efficiency_FWM_BS}, while operating with much lower power, continuous-wave optical pumps. Moreover, the compact geometry, scalable fabrication process, and versatility make them a promising resource for future applications.

\begin{center}
\begin{figure*}
\begin{center}
\includegraphics[width=0.8\linewidth]{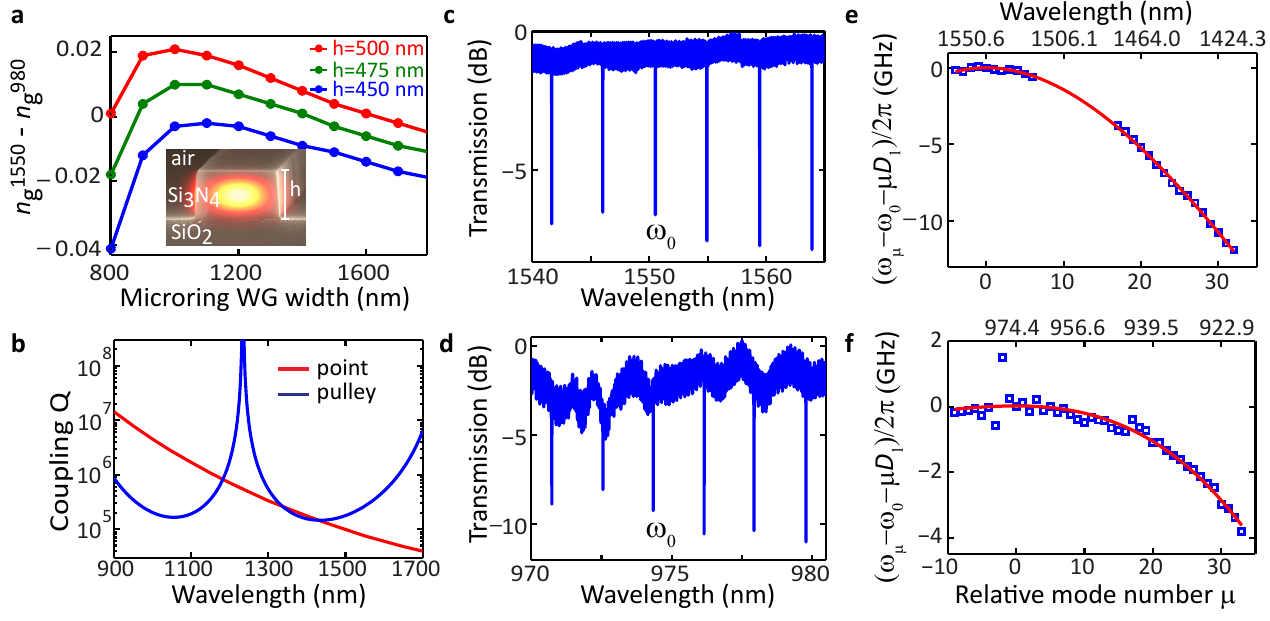}
\caption{\textbf{Device design, dispersion, and coupling characterization}. \textbf{a}, Simulated group index difference between the $1550$ nm and $980$ nm bands for various $\text{Si}_3\text{N}_4$ waveguide cross-sections with top air cladding and bottom oxide cladding (inset), from which a $1400$ nm $\times$ $480$ nm ring waveguide cross-section is chosen for this work. \textbf{b}, Simulated coupling $Q$ for a $820$ nm wide waveguide coupling to the microring with gap of $250$ nm, considering both point coupling and pulley coupling with a coupling length of $25\ \mu$m. \textbf{c-d}, Linear transmission scan of the fabricated sample for the $1550$ nm and $980$ nm bands, with intrinsic/total quality factors measured to be $\approx$~$4.5 \times 10^5/1.5 \times 10^5$ and $9.0\times 10^5/2.4\times 10^5$, respectively. \textbf{e-f}, Deviation of the measured resonance frequencies (markers) from an equidistant frequency grid $\omega_0 + \mu D_1$ for the $1550$ nm and $980$ nm bands, where $D_1^{1550}/2\pi = 572.39$ GHz $\pm\ 0.05$ GHz and $D_1^{980}/2\pi = 572.25$ GHz $\pm\ 0.05$ GHz (the uncertainty is due to the wavelength resolution and represents a one standard deviation value). The solid lines are the fitting curves using dispersion parameters $D_2/2\pi \approx -31.04$ MHz and $D_3/2\pi \approx 0.71$ MHz for the $1550$ nm band, and $D_2/2\pi \approx -3.43$ MHz and $D_3/2\pi \approx  - 0.30$ MHz for the $980$ nm band.}
\label{fig:Fig2}
\end{center}
\end{figure*}
\end{center}

\noindent \textbf{System Design} The silicon nitride (Si$_3$N$_4$) platform supports high $Q$ optical modes in the $1550$ nm and $980$ nm bands together with a relatively large Kerr nonlinearity, both of which are needed for efficient and low power frequency conversion. We use a $40$ $\mu$m radius microring resonator (Figs.~1a-b) to implement FWM-BS, with the fabrication details provided in the Methods. The first FWM-BS scheme under study is illustrated in Fig.~1c, where the interference of two $1550$ nm band pumps effectively creates a grating in the Si$_3$N$_4$ $\chi^{(3)}$ nonlinear medium that scatters the $980$ nm input signal to two idlers, which are separated from the signal by an amount equal to the difference in pump frequencies. These two idlers are labeled as $i-$ and $i+$ ($i+$ denotes the idler with higher frequency), similar to the Stokes and anti-Stokes sidebands in Raman scattering. In whispering-galley-mode geometries, phase matching is satisfied by the correct choice of azimuthal mode numbers, while energy conservation has to be met through careful engineering. The conversion efficiencies for $i\pm$ are a function of their frequency detunings with respect to their corresponding cavity modes, which can be derived as $\delta \hat{\omega}_{\pm} \approx (\hat{\omega}_{i\pm} - \hat{\omega}_s) \mp |\hat{\omega}_{p1}-\hat{\omega}_{p2}|$ (see Supplementary Section I.A), where $\hat{\omega}_{p1,p2,s,i\pm}$ denotes the resonance frequency accommodating the pump $1$, pump $2$, the signal $s$, and idler $i_{\pm}$, respectively. The resonance frequencies in each band can be approximated by the Taylor series $\hat{\omega}_\mu =\hat{\omega}_0 + D_1 \mu + \frac{1}{2} D_2 \mu^2 + \frac{1}{6} D_3 \mu^3...$, where $\hat{\omega}_0$ is the reference frequency, $D_1/2\pi$ is the free spectral range (FSR) of the resonator at $\hat{\omega}_0$, $D_{n}\ (n=2,,3...)$ are parameters characterizing the resonance dispersion, and $\mu$ is an integer representing the relative mode order number with respect to $\hat{\omega}_0$. As a result, the frequency detuning for $i\pm$ corresponding to pump separation of $|\mu|$ FSRs is given by $\delta \hat{\omega}_{\pm |\mu|} \approx \pm \delta D_{1} |\mu| + \frac{1}{2} \delta D_{2, \mp} \mu^2 ...$, where $\delta D_{1} \equiv D_{1}^{980} - D_{1}^{1550}$ and $\delta D_{2, \pm} \equiv D_{2}^{980} \pm D_{2}^{1550}$, with the $980$ and $1550$ superscripts marking the reference wavelengths in the $980$ nm and $1550$ nm bands, respectively. The first order frequency matching condition (i.e., $\delta D_{1} \approx 0 $) is satisfied by using a ring waveguide geometry that exhibits $n_{g}^{1550}=n_{g}^{980}$ (Fig.~2a), where $n_g$ stands for the group index and is inversely proportional to the FSR. Moreover, it is important to be in the overcoupled regime for the $980$ nm band to extract the generated idlers efficiently while maintaining a reasonably high $Q$ in the $1550$ nm band for resonance enhancement. The conventional point coupling, which relies upon the evanescent field overlap between the modes of a straight waveguide and the resonator, presents difficulty in doing so since the coupling becomes significantly weaker at shorter wavelengths (Fig.~2b). This problem is solved here by implementing the pulley coupling scheme~\cite{ref:Chin_Ho_res_wg_design,ref:Adibi_grp_pulley,ref:Spencer_Bowers_Si3N4_resonators}, which has an increased interaction length (Fig.~1b) and thus the coupling $Q$ depends on both the evanescent field overlap and phase matching. Because the resonant modes in the $1550$ nm band extend further outside the ring core than in the $980$ nm band, resulting in larger mode overlap, we choose an access waveguide width that is phase matched (mismatched) to the resonant modes in the $980$ nm ($1550$ nm) band, resulting in similar $Q$ values in the two bands. This is demonstrated by the simulation result in Fig.~2b, and confirmed by linear transmission measurements of the fabricated devices. By varying the coupling parameters (gap and coupling length) across multiple devices, we find a device operating in the overcoupled regime in both the $1550$ nm and $980$ nm bands, with loaded $Q$ values $\approx$~$1.5\times 10^5$ (linewidth $\kappa/2\pi \approx 1.29$ GHz) and $2.4 \times 10^5$ ($\kappa/2\pi \approx 1.28$ GHz), respectively (Fig.~2c-d). In addition, with the aid of a $200$ MHz resolution wavemeter, the resonances in each band can be measured (Fig.~2e-f), from which we extract $D_1^{1550}/2\pi = 572.39$ GHz $\pm\ 0.05$ GHz and $D_1^{980}/2\pi = 572.25$ GHz $\pm\ 0.05$ GHz, where the uncertainty is due to the wavemeter resolution and represents a one standard deviation value. The difference in FSR between the $1550$ nm and $980$ nm bands is well below the resonator linewidth, indicating the first order frequency matching condition has been satisfied. We also extract high-order dispersion terms $D_2$ and $D_3$ in the two wavelength bands of interest and find that the resonator shows normal dispersion ($D_2 <0$) in both bands. In general, $D_2$ and $D_3$ need to be taken into account for the frequency detuning calculation of idlers when the two pumps are widely separated, as the deviation of resonance frequencies from an equally spaced frequency grid (separated by one FSR) becomes comparable to the resonance linewidth for large $|\mu|$ values.

\begin{center}
\begin{figure*}
\begin{center}
\includegraphics[width=0.8\linewidth]{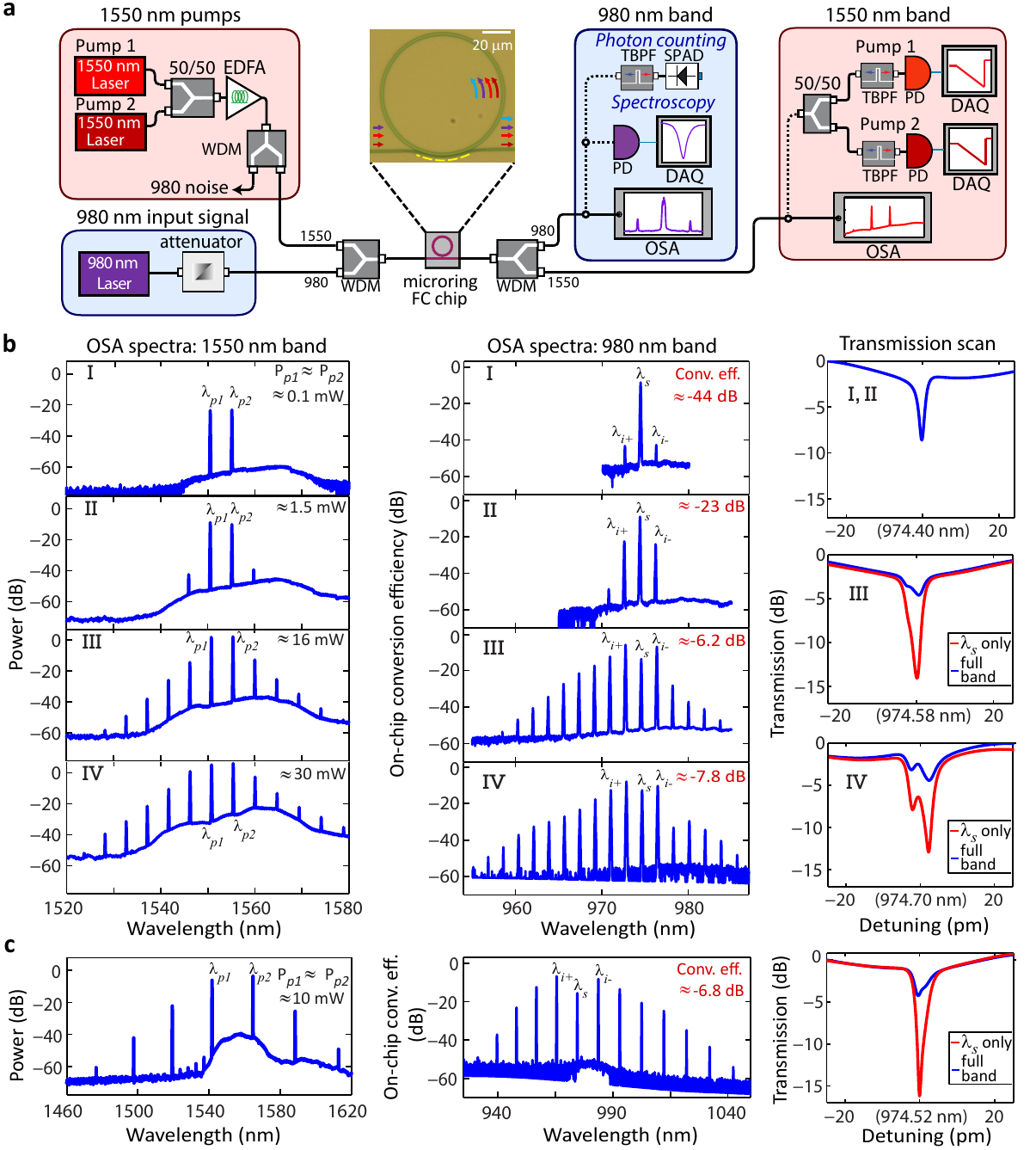}
\caption{\textbf{$980$ nm intraband frequency conversion}. \textbf{a}, Experimental setup for the $980$ nm intraband frequency conversion: EDFA, erbium-doped fiber amplifier; WDM: wavelength-division multiplexer; TBPF: tunable bandpass filter; SPAD: single-photon avalanche diode; PD: photodetector; DAQ: data acquisition; OSA: optical spectrum analyzer. \textbf{b}, Experimental frequency conversion results for two pumps separated by one free spectral range (FSR). The two pumps and the signal are located at $1550.6$ nm, $1555.2$ nm, and $974.4$ nm, respectively. Four cases (I-IV) corresponding to different pump power levels are displayed: the first two columns correspond to the OSA spectra in the $1550$ nm and $980$ nm band, respectively, and the last column shows the transmission scans of the signal around its thermally shifted resonance ($x$ axis origin). The transmission scan data is taken both with a narrowband filter centered on the signal resonance (rejecting all converted idlers), and across the full 980 nm band (no narrowband filter). In the transmission scan data for cases I and II, these two curves are identical. \textbf{c}, Experimental results for the two pumps separated by five FSRs. The two pumps are located at $1541.5$ nm and $1564.5$ nm and the signal is the same as the one FSR case ($974.4$ nm). In the 1550~nm band OSA spectra in \textbf{b}-\textbf{c}, a power of $0$ dB is referenced to $1$ mW.}
\label{fig:Fig3}
\end{center}
\end{figure*}
\end{center}
\noindent\textbf{$980$ nm intraband conversion.}
The experimental setup is shown in Fig.~3a, where the two $1550$ nm pumps are amplified and combined with the $980$ nm signal, and coupled to the chip using a lensed fiber (coupling loss $\approx$~$5.5$ dB and $6.5$ dB per facet for the $1550$ nm and $980$ nm band, respectively). To tune both pump lasers into resonance, narrowband filters (bandwidth $\approx 1$ nm) are placed in front of each $1550$ nm detector to select each pump while it scans across its resonance with decreasing frequency detuning, showing a characteristic triangular resonance shape from a combined Kerr-nonlinear and thermal resonance shift~\cite{ref:Carmon_Vahala_self_stability}. As we reduce the spectral distance between one pump and its nearby resonance, the intracavity power rises and the dominant thermal shift sends all the resonances towards lower frequencies, effectively increasing the spectral distance of the other pump laser with respect to its accommodating resonance. Therefore, it is necessary to tune the two pump lasers in an interactive and iterative way until both are thermally locked on their respective resonances.

Once the pumps are thermally locked, we characterize the frequency conversion process through two types of measurements. The first is to spectrally resolve the light exiting the chip using an optical spectrum analyzer (OSA). The second is to examine signal depletion and idler generation by swept wavelength spectroscopy of the signal resonance. Figure 3b shows the experimental results when the two pumps are separated by one FSR and of nearly equal power. When the pump power is low (first column, case I), only two small sidebands ($i+$ and $i-$) are observed in the $980$ nm spectrum (second column, case I); as the pump power increases (case II), secondary pump peaks appear in the $1550$ nm spectrum from pump mixing, which in turn generate higher-order idlers in the $980$ nm spectrum. The on-chip conversion efficiency, defined as the converted idler photon flux at the waveguide output divided by the input signal photon flux at the waveguide input, reaches a maximum of $-6.2$ dB for both $i+$ and $i-$ at $16$ mW power per pump (case III). In addition, multiple high-order idlers are generated, though the red idlers decrease more rapidly than the blue ones due to an accidental mode shift of the $978$ nm resonance (Fig.~2f, $\mu=-2$) caused by mode interactions (see Supplementary Section III). If we further increase the pump power (case IV), more sidebands are observed in both the $1550$ nm and $980$ nm bands, but the conversion efficiency for $i+$ and $i-$ slightly decreases, indicating there is a saturation level for a specific idler. The OSA data is complemented by transmission measurements in the $980$ nm band (third column). First, we use a narrowband filter (centered on the signal resonance with bandwidth $\approx 0.4$ nm) to reject all of the generated idlers and provide a measurement of the signal transmission. As the pump power increases, the signal resonance evolves from overcoupled (cases I and II) to critically coupled (case III) and then undercoupled (case IV), indicating an increased intrinsic loss that is attributed to signal depletion due to the frequency conversion process. In addition, mode splitting starts to appear as the pump power increases, a phenomenon expected when there is a strong coupling between the signal and adjacent idlers (see next section and also Supplementary Section I.A). Next, we remove the narrowband filter and directly detect light in the whole $980$ nm band. At low pump powers (cases I and II), the transmission scan is the same as the linear case, but when the conversion efficiency is high (cases III and IV), the extinction ratio becomes much smaller, suggesting a significant amount of idler generation when the signal is on-resonance. Moreover, mode splitting is also evident at sufficiently high powers (case IV). We can also increase the separation between the two pumps. When they are separated by $5$ FSRs (Fig.~3c), the optimum on-chip conversion efficiency for $i+$ and $i-$ is similar to the $1$ FSR case, indicating the phase matching condition is well satisfied. Thus, in this configuration, frequency conversion with a signal translation as low as $1.8$ nm (1 FSR) and as large as $9.0$ nm (5 FSR) can be achieved with an efficiency as high as $25~\%$ in each first-order idler.

\begin{center}
\begin{figure*}
\begin{center}
\includegraphics[width=0.8\linewidth]{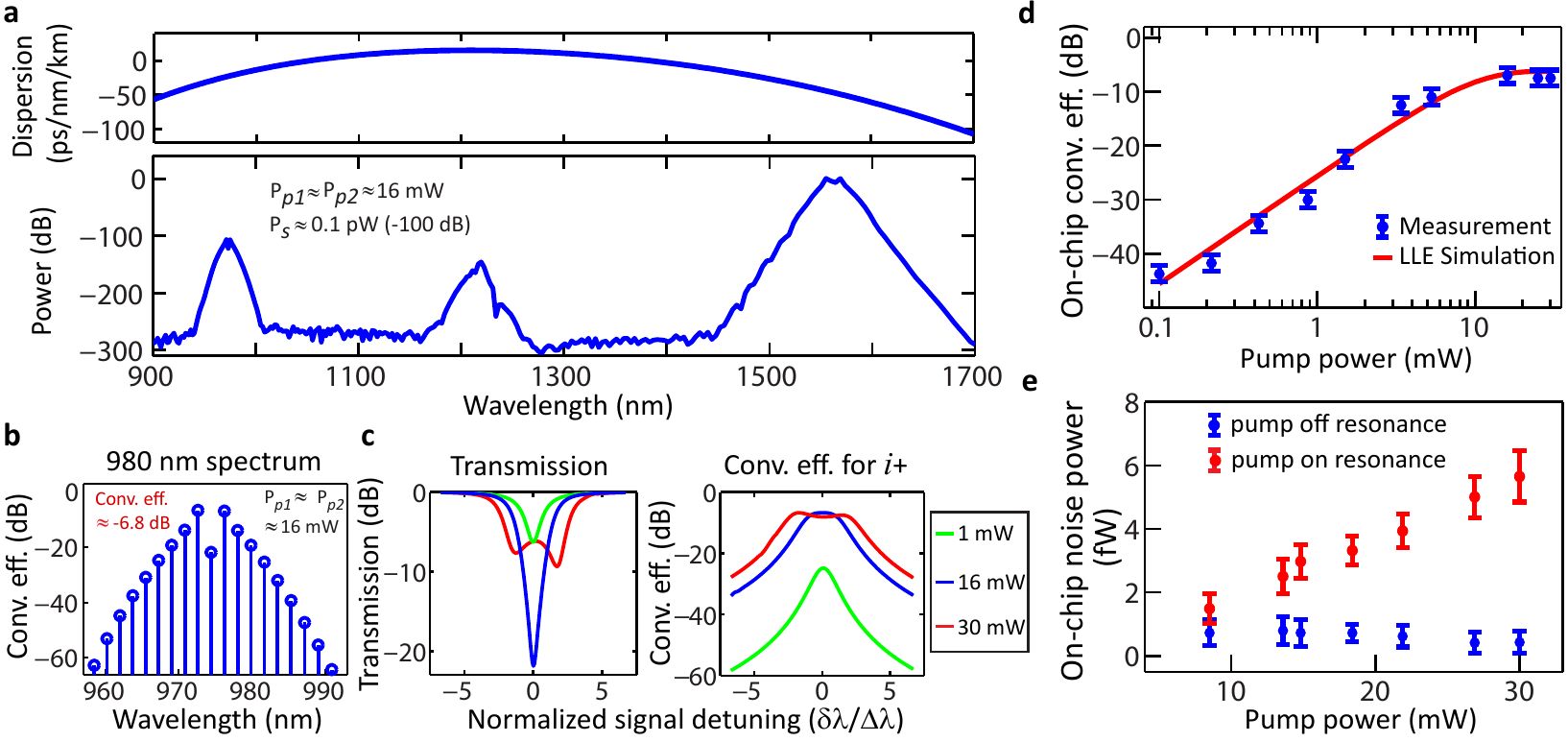}
\caption{\textbf{Numerical simulations and SNR measurement for 980 nm intraband conversion.} \textbf{a}, (Bottom) Simulated optical spectrum at the output of the waveguide based on a modified Lugiato-Lefever equation for the pump power of $16$ mW per pump and signal power of $0.1$ pW. (Top) Plot of the dispersion parameter, obtained from an eigenfrequency mode solver for the micresonator under study and converted to an equivalent group velocity dispersion. \textbf{b}, zoomed-in $980$ nm spectrum with the $y$ axis normalized to the photon flux of the input signal.  \textbf{c}, Simulated transmission scan of the signal resonance (left) and the on-chip conversion efficiency for $i+$ (right) for several different pump powers using the modified LLE method. The $x$ axis is the wavelength detuning normalized by the linewidth of the resonance ($2\Delta \lambda \approx 4$ pm). In both \textbf{a} and \textbf{c}, a power of $0$ dB is referenced to $1$ mW. \textbf{d} Measured (blue markers) and simulated (red line) conversion efficiency of $i+$ as a function of pump power. \textbf{e}, Measured on-chip background noise at $i+$ frequency with $\approx 120$ GHz detection bandwidth for various pump powers with the pumps off- and on-resonance with their respective cavity modes. The error bars in \textbf{d} and \textbf{e} are due to fluctuations in the detected signal by the SPAD, and represent a one standard deviation value. In all of the above, the $1550$ nm pump separation is $1$ FSR.}
\label{fig:Fig4}
\end{center}
\end{figure*}
\end{center}

\vspace{0.1in}

\noindent\textbf{Numerical simulations} Four-wave mixing in $\chi^{(3)}$ resonators can be described by the Lugiato-Lefever equation (LLE), a version of the nonlinear Schrodinger equation that includes driving and dissipation and which has been successfully applied to study microcavity frequency combs~\cite{ref:Coen_LLE}. To understand our experimental results, numerical simulations based on a modified form of the conventional LLE formalism are carried out. The modification is needed for FWM-BS to allow for three independent inputs (i.e., two pumps and one signal). In addition, coupled mode equations are also used~\cite{ref:Matsko_Maleki_hyperparametric,ref:Chembo_coupled_mode_combs}. Though the two approaches are essentially equivalent~\cite{ref:Hansson_CMT_LLE,ref:Chembo_Menyuk_LLE_coupled_mode}, they differ in detailed forms and numerical implementation, which may render one better than the other depending on the scope of the problem. For example, the coupled mode equations are usually preferred for revealing the underlying physics, while the LLE method is capable of capturing all relevant physical mechanisms by automatically including a complete set of modes into simulation.

Figure 4a shows the simulation results based on the modified LLE method (see Methods) for the case of $16$ mW power per pump. Compared to the corresponding experimental results (Fig.~3b case III), we find the modified LLE simulation successfully models the pump mixing in the $1550$ nm band and the multiple idler generation in the $980$ nm band (Fig.~4b). In addition, there is a small noise band in the anomalous dispersion region ($1100$ nm to $1300$ nm), which arises from modulation instability processes (i.e., one photon from the $980$ nm signal and one photon from the $1550$ nm pumps are converted to nearly degenerate photon pairs near the wavelength of $1200$ nm), though such processes are generally frequency-mismatched and small enough to be neglected. The transmission scan of the signal resonance from the LLE simulation (Fig.~4c, left) also shows good agreement with the experimental data, and the observed mode splitting is explained by a simplified set of coupled mode equations to result from strong couplings induced by FWM-BS between the signal and its adjacent idlers (Supplementary Section I.A). Generally, increasing pump powers leads to larger mode splitting and increased conversion bandwidth for the signal photon (Fig.~4c, right), but not necessarily higher conversion efficiency for a specific idler as its conversion efficiency will eventually saturate.

\vspace{0.1in}

\noindent \textbf{SNR measurement} Along with conversion efficiency, a key metric for quantum photonic applications (as well as some classical applications) is the background noise generated in the converted idler band. In general, one may expect that noise to originate due to the presence of strong pump fields, which through processes such as Raman scattering, may generate photons that are either directly resonant with the converted idler, or are resonant with the input signal and get frequency converted together with it. One advantage of the scheme presented here is that the signal and idler are about $600$ nm away from the strong pump fields (and are on their anti-Stokes side), suggesting that noise due to Raman scattering should be limited.

\begin{center}
\begin{figure*}
\begin{center}
\includegraphics[width=0.8\linewidth]{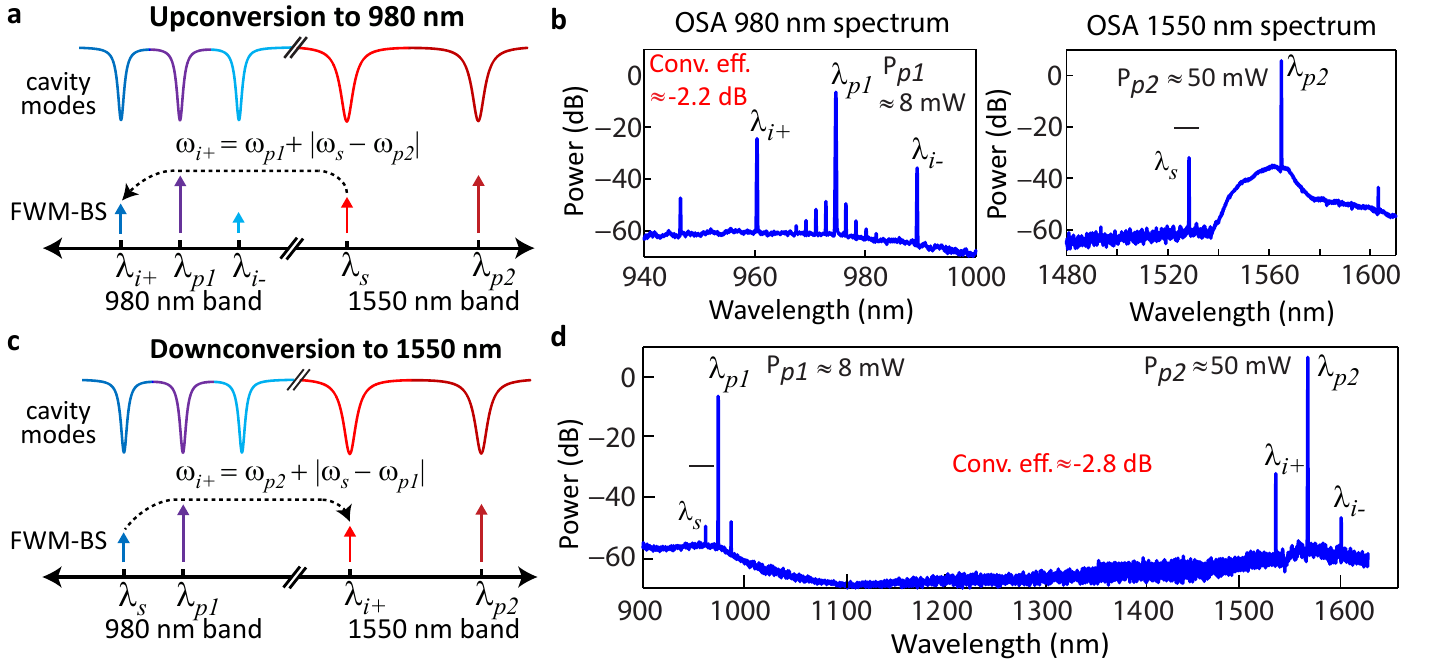}
\caption{\textbf{Wideband frequency conversion interface to the telecommunications band}. \textbf{a}, Schematic of the signal upconversion from the $1550$ nm band to the $980$ nm band. \textbf{b}, Representative experimental result for upconversion: the two pumps and the signal are located at $974.4$ nm, $1564.5$ nm and $1528.0$ nm, respectively. The small horizontal bar above the signal indicates its input power. In the $980$ nm spectrum, other than $i+$ and $i-$ and a group of small idlers near the $980$ pump which correspond to the upconverted ASE noise of the EDFA, there is another small peak near $946.2$ nm due to the mixing between $i+$ and the $980$ nm pump. Similarly, in the $1550$ nm spectrum the small peak near $1602.8$ nm is due to the the mixing of the signal and the $1550$ nm pump.  \textbf{c}, Schematic of the signal downconversion from the $980$ nm band to the $1550$ nm band. \textbf{d}, Representative experimental result of downconversion showing a full spectrum: the two pumps and the signal are located at $974.4$ nm, $1559.8$ nm, and $961.9$ nm, respectively. The small horizontal bar above the signal indicates its input power. In \textbf{b} and \textbf{d}, a power of $0$ dB is referenced to $1$ mW.}
\label{fig:Fig5}
\end{center}
\end{figure*}
\end{center}

To determine the suitability of this system as a QFC interface, we consider the conversion efficiency and SNR for an input signal at the single photon level ($\approx 1$~pW, corresponding to photon flux $\approx 5 \times 10^6$ s$^{-1}$, which is consistent with the photon flux produced by systems such as a quantum dot single photon source~\cite{ref:Michler_book_2009} triggered at $50$ MHz and with a source efficiency of $10~\%$). A single-photon avalanche diode (SPAD) is used to measure the power in the converted idler band, which is spectrally isolated from the $1550$ nm pumps using a WDM and a $0.4$ nm bandwidth bandpass filter (Fig.~3a). The on-chip conversion efficiency, background noise power, and SNR are determined from knowledge of the coupling loss and insertion losses of the filtering elements. As shown in Fig.~4d, the on-chip conversion efficiency as a function of pump power for the $i+$ idler agrees with the LLE simulation, and peaks at $\approx 25~\%$. We next measure the noise power by turning off the input signal and considering two cases, when the pumps are on-resonance with their respective cavity modes and when they are off-resonance (Fig.~4e). Off resonance, any measured noise is found to be broadband emission from the erbium-doped fiber amplifier (EDFA) used for the pump, and can be significantly reduced by filtering at the input side, prior to the chip (see Fig.~3a, WDM filter placed after the EDFA). On resonance, we observe noise above the EDFA-determined background, at the level of a few fW. This noise is mainly contributed by the Si$_3$N$_4$ microresonator, though its origin is still under investigation (see Supplementary Section VI). At the optimum conversion efficiency ($25~\%$ at pump power $\approx 16$ mW), the SNR is larger than $80:1$ for the aforementioned input signal photon flux $\approx 5 \times 10^6$ s$^{-1}$. Moreover, for triggered sources, we only need to consider generated noise that temporally coincides with the single photon emission, which is produced at well defined times (the triggering times) and over well defined intervals (the spontaneous emission lifetime; typically $\approx$~1~ns for an InAs/GaAs quantum dot~\cite{ref:Michler_book_2009}). Therefore, either gated detection or temporal gating of the signal via amplitude modulation~\cite{ref:Ates_Srinivasan_Sci_Rep} can be used, with a typical duty cycle of $5~\%$ (for a $50$ MHz triggering rate). Since the measured background noise comes from a continuous-wave pump, we expect the corresponding noise level in the gated case to be reduced by a factor equal to the duty cycle, which will push the SNR to $>1000:1$ at the optimum conversion efficiency.

\begin{center}
\begin{figure*}
\begin{center}
\includegraphics[width=0.8\linewidth]{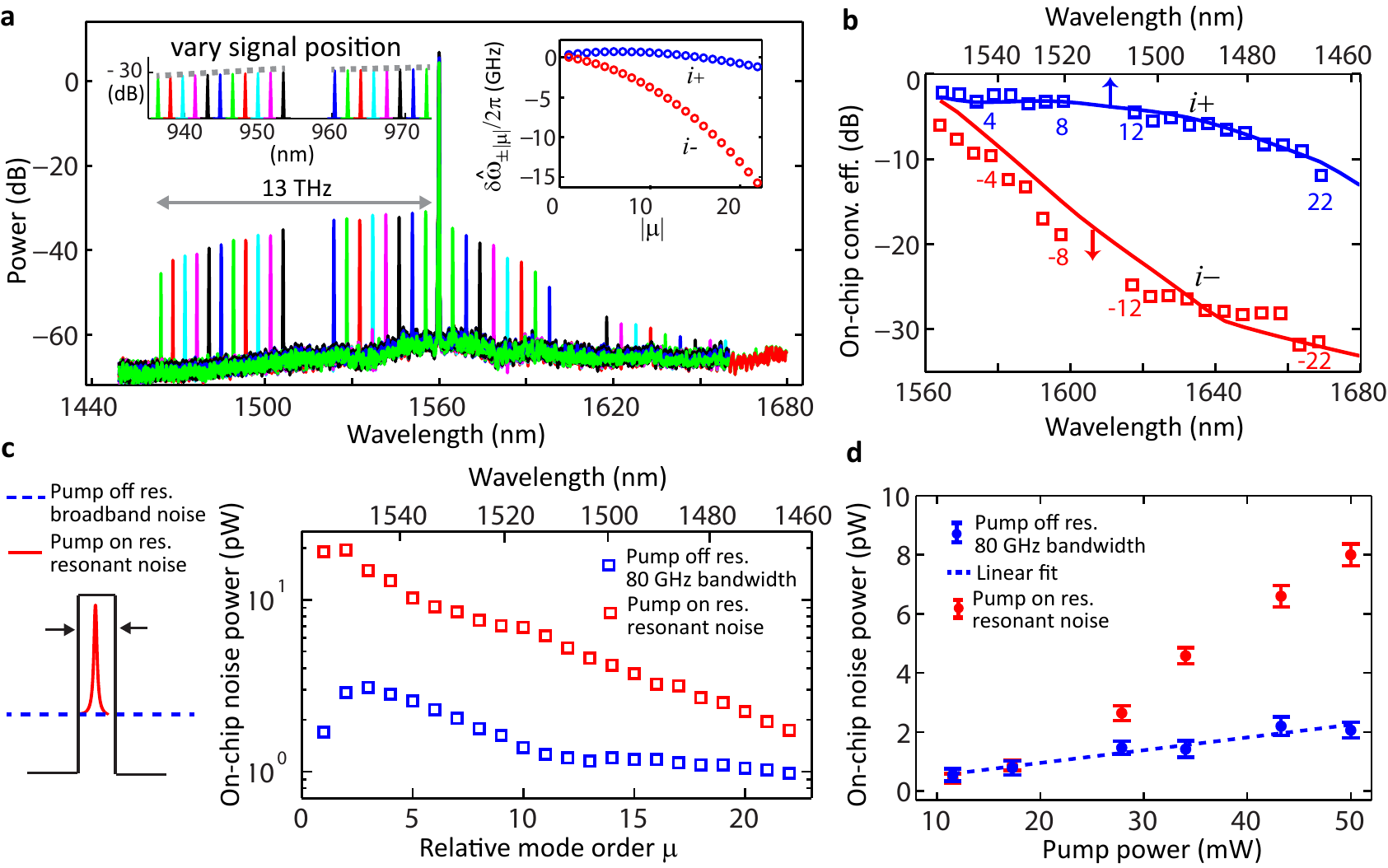}
\caption{\textbf{Wideband frequency downconversion bandwidth and background noise measurement.}. \textbf{a} Superimposed OSA spectra in the $1550$ nm band for different signal separation from the $980$ nm pump with fixed pump powers in the $1550$ nm and $980$nm bands ($50$ mW and $8$ mW, respectively). The inset on the left side illustrates the position of the signal input from $980$ nm band lasers covering wavelength ranges of $930$ nm to $955$ nm and $960$ nm to $990$ nm. In addition, the input signal power decreases as it goes to shorter wavelengths, which also contributes to the reduced idler power as observed in the $1550$ nm spectrum. For both the main figure and its left inset, a power of $0$ dB is referenced to $1$ mW. The inset on the right side shows the calculated frequency detuning for $i+$ and $i-$ at different idler positions. \textbf{b} Measured (markers) and simulated (solid lines) conversion efficiencies for $i+$ and $i-$ as a function of the signal separation from the $980$ nm pump. \textbf{c}-\textbf{d}, Measured on-chip background noise power with a $80$ GHz detection bandwidth for the pump-off-resonance and the resonant noise for $i+$ when the pump is on-resonance for \textbf{c}: different idler positions while keeping the $1550$ nm pump power fixed at $50$ mW and for \textbf{d}: fixed idler position ($\mu=7$) while varying the $1550$ nm pump power. The error bars in \textbf{d} are due to fluctuations in the SPAD detection rate, and represent a one standard deviation value. The error bars in \textbf{c} are smaller than the marker size.}
\label{fig:Fig6}
\end{center}
\end{figure*}
\end{center}

\noindent\textbf{Telecommunications band frequency conversion interface} In addition to frequency conversion within the $980$ nm band, we can use the same device for frequency conversion between the $1550$ nm and $980$ nm bands. As illustrated by Figs.~5a and 5c, with one pump in each band, the signal can be upconverted (downconverted) from the $1550$ nm ($980$ nm) band to the $980$ nm ($1550$ nm) band. Although the pump power available from our $980$ nm laser is limited (maximum $\approx 8$ mW), it can be compensated by using a relatively strong pump in the $1550$ nm band, since the conversion efficiency is determined by the product of the two pump powers~\cite{ref:Uesaka_Kazovksy}. Figure 5b shows one example of upconversion, measured using the experimental setup shown in Fig.~S6 of the Supplementary, where the signal is separated from the $1550$ nm pump by $8$ FSRs (i.e., $\mu=8$). In the $980$ nm band spectrum, the converted blue idler ($i+$) shows a conversion efficiency $\approx -2.2$ dB while the red idler ($i-$) is much weaker (as in the $980$ nm intraband case, we define conversion efficiency in terms of photon flux). The difference in the conversion efficiency indicates different frequency detunings for the two idlers, which is expected as high-order dispersion terms become important when $|\mu|$ is large (see Supplementary Section IV for more detailed discussions). In addition, we find the amplified spontaneous emission (ASE) around the $1550$ nm pump (largely due to the EDFA) is also upconverted to the $980$ nm band (small peaks around the $980$ nm pump). By adding a narrowband filter after the EDFA, the ASE of the $1550$ nm pump can be suppressed.

As one would expect, if efficient $1550$ nm to $980$ nm upconversion can be achieved in this system, efficient $980$ nm to $1550$ nm downconversion should also be achievable, and easily measured by simply moving the input signal to the $980$ nm band (Fig.~5c). Figure 5d shows a representative downconversion spectrum when the separation between the input signal and the $980$ nm band pump is $7$ FSRs ($\mu=7$), where in this case a narrowband ASE rejection filter surrounding the $1550$ nm pump is applied (see Supplementary Fig.~S7a). Along with a conversion efficiency for $i+$ that is similar to the upconversion case ($\approx -2.8$ dB), we see the ASE rejection filter has successfully suppressed the spurious peaks present in the upconversion spectrum.

To further explore the properties of wideband frequency conversion based on the FWM-BS scheme, we focus on the downconversion case by varying the signal position in the $980$ nm band while keeping pump powers in the $1550$ nm and $980$ nm band $\approx$~$50$ mW and $8$ mW, respectively (Fig.~6a). As we increase the separation between the signal and the $980$ nm pump (i.e., $\mu$ increases), the blue idler $i+$ moves away from the $1550$ nm pump by an equal amount of frequency shift with slowly decreasing power. By contrast, the red idler $i-$ shows slightly worse conversion efficiency than $i+$ when $\mu$ is small, but as $\mu$ increases its power decreases much more rapidly. Their different performance can be understood by calculating the corresponding frequency detunings (Fig.~6a inset, right), which reveal that $i+$ has a small frequency detuning for up to $22$ FSR separation from the $1550$ pump, while the frequency detuning for $i-$ is already comparable to the resonance linewidth after $5$ FSRs. In Fig.~6b, the conversion efficiencies for both $i+$ and $i-$ are extracted from Fig.~6a (markers), and agree with the simulation results based on coupled mode equations (solid lines, see Supplementary Section I.B). With less than $60$ mW total pump power, the maximum conversion efficiency for $i+$ is measured to be $\approx$~$60~\%$ ($-2.2$ dB), and the spectral range over which the input signal can be tuned without reducing the conversion efficiency by more than $3$~dB (i.e., conversion efficiency $>30~\%$) is estimated to be nearly $8.2$ THz. Even when the signal is $13$ THz away from the $980$ nm pump, the conversion efficiency for $i+$ is still more than $10~\%$.

To measure the SNR level for the wideband frequency conversion (see Supplementary Section V), we turn off the signal and measure the noise at the idler frequency while keeping the pumps on. Narrowband filters are added at the input side for each pump to suppress its ASE level so that the single-photon-level signal and idlers are not overwhelmed by the ASE noise. For the upconversion case as shown in Fig.~5a, the $1550$ nm pump is expected to generate little noise in the $980$ nm band as already demonstrated in the $980$~nm intraband frequency conversion (Fig.~4e), although it is not trivial to suppress the $980$ nm pump to sub pW level on the detection side ($>100$ dB suppression) so that single-photon-level noise can be measured. For the downconversion case, the noise is found to be mostly from the $1550$ nm pump and the contribution of the $980$ nm pump is negligible. At the output, after filtering out the $1550$ nm pump, the transmitted light goes through a narrowband filter with a tunable bandwidth ($32$ pm to $600$ pm) and center wavelength ($1460$ nm to $1560$ nm) before detection by a SPAD. When the $1550$ nm pump is off-resonance, the noise power scales with the filtering bandwidth and exhibits a smooth spectral response as we shift the center wavelength of the filter, indicating it is a broadband noise. On the other hand, when we tune the pump into resonance, the detected noise initially stays the same if the narrowband filter is out of resonance, but increases significantly if the filter is centered on the idler resonance. The increased noise power changes little as we adjust the filtering bandwidth, suggesting this portion of noise is resonant. Figure 6c shows the measured on-chip noise power for idlers with different spectral separation from the $1550$ nm pump (power fixed $\approx 50$ mW). The pump-off-resonance noise is identified as anti-Stokes Raman scattering generated inside fiber after the EDFA, as the noise is almost the same if we remove the chip and introduce a similar insertion loss between two lensed fibers (see Supplementary Section VI for more details). On the other hand, the resonant noise (i.e., noise generated at the idler frequency with bandwidth limited by the cavity linewidth) for the pump-on-resonance case is believed to be from the $\text{Si}_3 \text{N}_4$ microresonator, and decreases rapidly as the idler moves away from the pump. For example, the resonant noise is measured to be $\approx$~$8$ pW for the $\mu=7$ idler ($1528$ nm resonance). Assuming the input signal is from a single photon emitter with a photon flux $\approx 5\times 10^6$ s$^{-1}$ ($\approx 1$ pW power), the converted blue idler has a power $\approx$~$0.52$ pW ($\approx -2.8$ dB conversion efficiency). At first glance, it seems the signal will be overwhelmed by the noise. However, just as we discussed in the intraband conversion case, for a triggered single photon source, either gated detection or temporal gating of the light exiting the frequency conversion device can be used, with a typical duty cycle of $5~\%$. This will improve the SNR to be more than unity under current conditions. Another straightforward route to reducing the noise power is to design a frequency-matched idler that has a larger spectral separation from the pump. Improving the Si$_3$N$_4$ material quality may also be important, as the resonant noise is suspected to be correlated with the significant material absorption inferred by a relatively large thermo-optic shift for the pump modes in the $1550$ nm band (see Supplementary Section III). Figure 6d shows that for a fixed idler ($\mu=7$ at $1528$ nm), the pump-off-resonance noise has a linear dependence on the pump power while the resonant noise is clearly nonlinear. Since the conversion efficiency is determined by the product of two pump powers, and the $980$ nm pump is expected to generate little noise in the $1550$ nm band due to the large spectral separation, it seems possible to improve the SNR dramatically by using a relatively large $980$ nm pump (e.g., $40$ mW) and a smaller $1550$ nm pump (e.g., $10$ mW). Optimizing the noise performance of this downconversion interface will be a subject of future studies.

\vspace{0.1in}

\noindent \textbf{Discussion} We have demonstrated efficient single-photon-level frequency conversion through four-wave-mixing Bragg scattering (FWM-BS) in monolithic Si$_3$N$_4$ microring resonators.  In contrast to parametric amplification (PA), which is the basis of the majority of prior work in Kerr-based nanophotonic devices~\cite{ref:Moss_Gaeta_Lison_NLO}, FWM-BS avoids spontaneous emission noise and can be, in principle, noiseless~\cite{ref:McKinstrie_FWM_BS}.  We have demonstrated background noise powers in the fW to pW range, and potential routes to reduce this further have been identified.  The performance of these devices is already comparable to what has been demonstrated for FWM-BS in highly nonlinear fibers and photonic crystal fibers and in some respects, such as pump power required and extent of the spectral translation range, exceeds them.  Moreover, the versatility of the FWM-BS process has been demonstrated.  Because the frequency translation range is set by the difference in pump frequencies (in contrast to frequency conversion by either PA or the $\chi^{(2)}$ nonlinearity), our devices enable both intraband conversion (translation range between 1.8~nm and 9.0~nm shown with no degradation in performance) and interband conversion (upconversion and downconversion over a translation range $>$560~nm).  These devices are already suitable for proof-of-principle QFC experiments with single photon Fock states, and the concepts we utilize - dispersion engineering to ensure frequency matching and tailored waveguide coupling to ensure efficient injection of the signal and pumps and extraction of the converted idler - can be readily applied to connect visible-wavelength quantum light sources (e.g., based on color centers in diamond or trapped atoms and ions) with the telecommunications band.

There are several topics to be addressed in the pursuit of such goals.  The use of a resonantly-enhanced process comes at the expense of bandwidth, which is set by the cavity linewidth $\kappa/2\pi \approx 1.3$~GHz in both the 980~nm and 1550~nm bands.  Importantly, this bandwidth is large enough to accommodate the photons generated by single InAs/GaAs quantum dots~\cite{ref:Michler_book_2009}, our intended target application, and recently demonstrated quantum light sources in the Si$_3$N$_4$ platform would also be compatible~\cite{ref:Dutt_Lipson_squeezing,ref:Ramelow_Gaeta_SiN_pairs}. The GHz bandwidth stands in contrast to compact wavelength converters based on cavity optomechanics~\cite{ref:Dong_Wang_OM_dark_mode,ref:Hill_Painter_WLC,ref:Liu_yuxiang_wlc}, which have been limited to the MHz range (at most). Nevertheless, even with a sufficiently large bandwidth, the precise temporal profile of the incoming photon's wavepacket is an important consideration when determining the efficiency with which it can be coupled into the frequency conversion cavity~\cite{ref:Bader_Leuchs_cavity_loading,ref:Liu_Du_single_photon_loading}.  More generally, bandwidth and temporal shaping to efficiently load the cavity are topics that will need to be considered when moving from frequency conversion of continuous-wave signals to pulsed light. Finally, while the focus of our discussion has been on the demanding application of QFC, the potential classical applications of this work are also significant.  This includes add-drop wavelength multiplexing and cross-connect switching that utilize the interband FWM-BS process, and spectral translation of signals across hundreds of nanometers to connect, for example, atomic frequency references with telecommunications-band signals.

\vspace{3mm} \noindent \large{\textbf{Methods}}

\noindent \small{\textbf{Device fabrication}}
First, the layer stack is created by low-pressure chemical vapor deposition of a 480~nm thick Si$_3$N$_4$ layer on top of a 3~$\mu$m thick SiO$_2$ layer, which in turn was grown via thermal oxidation of a 100~mm Si wafer.  The wavelength-dependent refractive index and thickness of the layers are determined using a spectroscopic ellipsometer, with the data fit to an extended Sellmeier model.  After dicing into chips, the microring-waveguide devices are created by electron-beam lithography of a negative tone resist, followed by reactive ion etching of the Si$_3$N$_4$ using a CF$_{4}$/CHF$_3$ chemistry, removal of deposited polymer and remnant resist, and annealing at 1150~$^{\circ}$C in an O$_{2}$ environment for 3 hours.

\noindent \small{\textbf{Numerical simulation}}
The simulations presented in this work are based on coupled mode equations as well as a modified LLE method. For the $980$ nm intraband conversion, we first study a simplified set of coupled mode equations (which neglects pump mixing in the $1550$ nm band and high-order idler generation in the $980$ nm band) to gain some physical insight (Supplementary Section I.A.), from which the mode splitting observed in the experiment (Fig.~3b, case IV) can be easily understood. Next, a modified LLE method is developed (Supplementary Section I.C), which is essentially a systematic way to implement the same coupled mode equations by including a complete set of resonant modes. The simulation results shown in Figs.~4a and 4c show good agreement between the LLE simulation and the experimental results, indicating it has captured all relevant physical mechanisms. As for the wideband frequency conversion between the $1550$ nm and $980$ nm bands, since the pump mixing is absent, the LLE method is expected to generate almost the same results as the coupled mode equations. We use the coupled mode equations to study the device's performance, as its numerical computation is much faster than the LLE method (Supplementary I.B). Additional discussions on the upconversion and the effect of pump detunings to the frequency conversion process are also provided in the Supplementary Material (Section IV).

\noindent \small{\textbf{Frequency conversion metrics}}
We characterize our frequency conversion devices based on their on-chip conversion efficiency, noise power, signal-to-noise ratio for a single-photon-level input, and required pump power. The on-chip conversion efficiency is defined as the ratio of a given idler's photon flux at the coupling waveguide output to the signal photon flux at the coupling waveguide input.  The background noise power is defined as the power at a given idler's spectral position taken over a specified bandwidth (determined by a narrowband filter) and with the input signal switched off and both pumps turned on.  The SNR for single-photon-level inputs is estimated by attenuating the input signal level to a photon flux consistent with that achieved by quantum dot single photon sources (e.g., 1~pW~$\approx5{\times}10^{6}$~s$^{-1}$), and measuring the photon flux in a given idler band with both the signal on and off. Finally, the pump powers listed are on-chip and at the waveguide input.
%\bibliographystyle{OSA}
%\bibliography{KS_FWM_BS}

\noindent \textbf{Acknowledgements} Q.L. acknowledges
support under the Cooperative Research Agreement between the
University of Maryland and NIST-CNST, Award 70NANB10H193. The authors thank Lawrence Van Der Vegt from Yenista Optics for the loan of a $1550$ nm tunable filter, and Scott Papp from NIST Boulder for helpful comments.

\newpage
\onecolumngrid \bigskip

\begin{center} {{\bf \large SUPPLEMENTARY MATERIAL}}\end{center}

\setcounter{figure}{0}
\makeatletter
\renewcommand{\thefigure}{S\@arabic\c@figure}

\setcounter{equation}{0}
\makeatletter
\renewcommand{\theequation}{S\@arabic\c@equation}

\section{Theory and numerical simulations}
\vspace{-0.1in}
\begin{figure}[htb]
  \begin{minipage}[c]{0.4\textwidth}
  \centering
{\includegraphics[width=\linewidth]{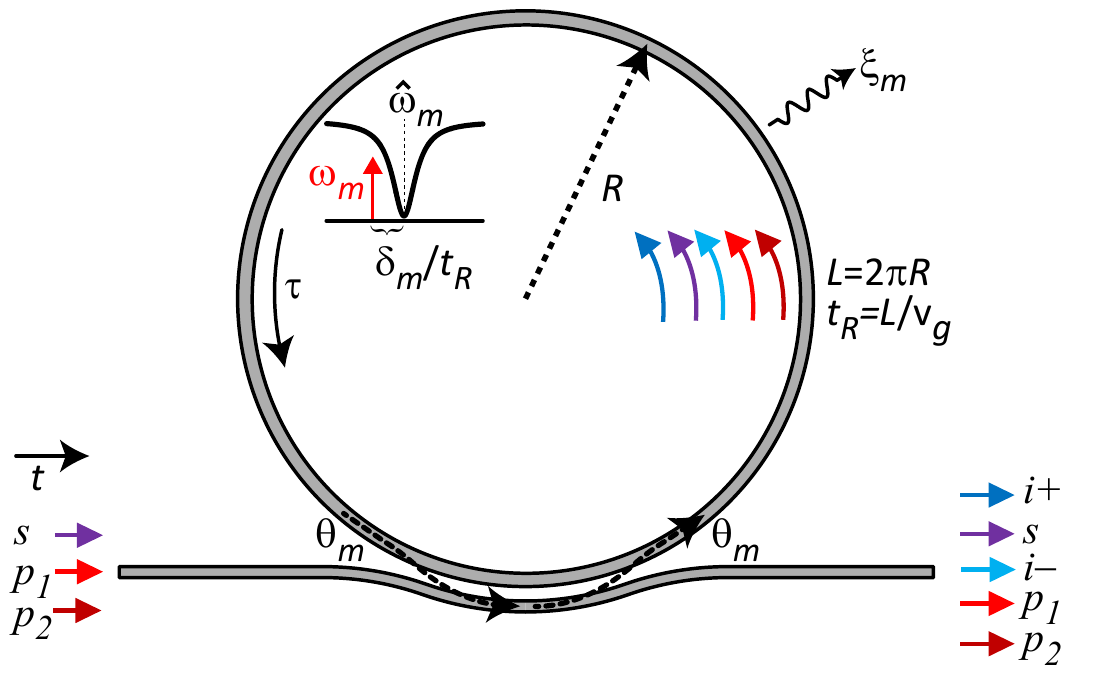}
}
  \end{minipage}
  \begin{minipage}[c]{0.5\textwidth}
  \centering
  \begin{tabular}{|c|c|c|c|}
\hline
Variable name & Description & Equation \\ [0.5ex]
\hline
$s$     & input signal           & - \\
$p1$    & pump 1                 & - \\
$p2$    & pump 2                 & - \\
$i+$    & blue-shifted idler     & - \\
$i-$    & red-shifted idler      & - \\
$\tau$  & time                   & - \\
$t$     & slow time scale set by cavity & - \\
$t_{R}$ & cavity round-trip time & - \\
$L$ & cavity round-trip length & - \\
$\omega_{m}$ & frequency of driving field near mode $m$ & - \\
$\hat{\omega}_{m}$ & frequency of mode $m$ & - \\
$E_{m}$ & intracavity electric field for mode $m$ & - \\
$E_{in,\omega_{m}}$ & driving field at frequency $\omega_{m}$ & - \\
$Q_{i,m}$ & intrinsic $Q$ for mode $m$ & - \\
$Q_{c,m}$ & coupling $Q$ for mode $m$ & - \\
$Q_{L,m}$ & total loaded $Q$ for mode $m$ & $Q_{L,m}^{-1}=Q_{i,m}^{-1}+Q_{c,m}^{-1}$ \\
$\alpha_{m}$ & cavity mode $m$ total loss rate & $\hat{\omega}_{m}t_{R}/(2Q_{L,m})$ \\
$\xi_{m}$ & intrinsic loss rate for mode $m$ & $\hat{\omega}_{m}t_{R}/Q_{i,m}$ \\
$\theta_{m}$ & waveguide power coupling rate for mode $m$ & $\hat{\omega}_{m}t_{R}/Q_{c,m}$ \\
$\delta_{m}$ & laser-cavity mode $m$ detuning& $\delta_{m}=(\hat{\omega}_{m}-\omega_{m})t_{R}$ \\
$\gamma_{m}$ & nonlinear coefficient for mode $m$ & see Section I.D  \\
\hline
\end{tabular}
  \end{minipage}
  \caption{(Left) Schematic and (Right) table listing the parameters used in the LLE and coupled mode equations models of the four-wave-mixing Bragg scattering process.}\label{Fig_S1}
\end{figure}

The conventional LLE formalism describes the Kerr nonlinearity in a $\chi^{(3)}$ microring resonator as ~\cite{ref:Coen_LLE_SM, ref:Chembo_coupled_mode_combs_SM}
\begin{equation}
t_R \frac{\partial E(t,\tau)}{\partial t} = \left [ - \alpha - i \delta_0 + iL \sum_{k \geq 2 }\frac{\beta_k}{k!} \left(i \frac{\partial}{\partial \tau}\right )^k + i \gamma L |E(t,\tau)|^2 \right ] E(t,\tau) + i\sqrt{\theta}E_{\text{in}},
\label{Eq_LLE}
\end{equation}
where $t_R$ is the round-trip time, $E(t,\tau)$ is the intracavity mean field with $|E(t,\tau)|^2$ representing the average power traveling inside the cavity ($\tau$ is the time and $t$ is a parameter measuring the slow time of the cavity), $\alpha$ and $\delta_0$ are the total cavity loss and detuning at the resonance frequency $\hat{\omega}_0$, respectively ($\alpha = \hat{\omega}_0 t_R/(2Q_L)$ with $Q_L$ being the loaded $Q$ and $\delta_0= (\hat{\omega}_0 - \omega_0)t_R$ with $\omega_0$ being the frequency of the driving field), $L$ is the round-trip length of the resonator, $\beta_k (k\geq 2)$ are the second and higher-order dispersion parameters of the ring waveguide, $\gamma$ is the Kerr nonlinear coefficient, $\theta$ is the power coupling coefficient between the resonator and the access waveguide or fiber ($\theta=\hat{\omega}_0 t_R/Q_c$ with $Q_c$ being the coupling $Q$), and $E_\text{in}$ is a continuous-wave driving field (input power $P_{\text{in}} = |E_\text{in}|^2$). Figure \ref{Fig_S1} schematically depicts the system under investigation, and includes a table of the aforementioned variables.

As mentioned in the main paper, the LLE formalism is equivalent to coupled mode equations ~\cite{ref:Hansson_CMT_LLE_SM, ref:Chembo_Menyuk_LLE_coupled_mode_SM}. In fact, $E(t,\tau)$ can be viewed as a collection of resonant modes $\{E_m(t)\}$ ($m$ is the azimuthal order) oscillating on an equally spaced frequency grid $\{\omega_m\}$ ($E(t,\tau)=\sum_m E_m(t)\exp(-i\omega_m \tau)$), whose governing coupled mode equation can be obtained through a Fourier transform of Eq.~\ref{Eq_LLE}. However, for the FWM-BS process we have three driving fields (instead of only one pump in the standard LLE formalism), and each can have its own detuning with respect to the nearest cavity mode. Thus, the coupled mode equations should present the following form:
\begin{equation}
t_R \frac{d E_m}{dt}= -(\alpha_m + i\delta_m) E_m  + i \gamma_m L \mathcal{F}\left \{ |E(t,\tau)|^2 E(t,\tau) \right \}_m + i\sqrt{\theta_m} E_{\text{in},m},
\label{Eq_CMT}
\end{equation}
where $\mathcal{F}$ stands for the Fourier transform, and all the parameters and expressions with subscript $m$ are meant to be computed at frequency $\omega_m$. In particular, the detuning parameter $\delta_m$ is given by $\delta_m=(\hat{\omega}_m - \omega_m)t_R$, where the resonance frequency $\hat{\omega}_m$ is determined by a self-consistent equation $\hat{\omega}_m = \hat{\omega}_0 + \sum_{k\geq 1} L\beta_k (\hat{\omega}_m -\hat{\omega}_0)^k/(k!t_R)$. In the experiment, the frequencies of the driving fields for the two pumps and the signal ($\omega_{p1,p2,s}$) are free parameters that can be varied to achieve an optimum frequency conversion efficiency. On the other hand, the frequency of the generated idlers ($\omega_{i\pm}$) is determined by the energy conservation condition, and therefore is a function of $\omega_{p1,p2,s}$ (see Fig.~1c for the $980$ nm intraband conversion and Fig.~5 for wideband conversion in the main paper).

In the following, we will discuss the detailed implementations of coupled mode equations as well as a modified LLE method for the FWM-BS process.

\subsection{Coupled mode equations for $980$ nm intraband conversion}
 We start with a simplified set of coupled mode equations for the intraband frequency conversion in the $980$ nm band, that is, we neglect pump mixing in the $1550$ nm band and high-order idlers in the $980$ nm band. As a result, the modes under consideration are: two pumps ($p1$ and $p2$), the signal ($s$), and two idlers ($i+$ and $i-$). Following Eq.~\ref{Eq_CMT}, the coupled mode equation for pump $1$ can be written as
\begin{equation}
t_R \frac{d E_{p1}}{dt}=-(\alpha_{p1} + i \Delta \phi_{p1}) E_{p1} + i\sqrt{\theta_{p1} P_{p1}},
\label{Eq_CMT_p1}
\end{equation}
where we have combined $\delta_{p1}$ and the Kerr nonlinear shift into an effective detuning $\Delta \phi_{p1}$, which is given by
\begin{equation}
\Delta \phi_{p1} = (\hat{\omega}_{p1} - \omega_{p1})t_R -\gamma_{p1}L (|E_{p1}|^2 + 2|E_{p2}|^2).
\label{Eq_phip1}
\end{equation}
For a steady state, Eq.~\ref{Eq_CMT_p1} gives $E_{p1} = i \sqrt{\theta_{p1} P_{p1}}/(\alpha_{p1} + i\Delta \phi_{p1})$. Similarly, for pump $2$ we obtain $E_{p2} = i \sqrt{\theta_{p2} P_{p2}}/(\alpha_{p2}+ i\Delta \phi_{p2})$ and its effective detuning $\Delta \phi_{p2}$ as
\begin{equation}
\Delta \phi_{p2} = (\hat{\omega}_{p2} - \omega_{p2})t_R -\gamma_{p2}L (2|E_{p1}|^2 + |E_{p2}|^2).
\label{Eq_phip2}
\end{equation}

Next, we derive the coupled mode equations for the signal and idlers as
\begin{gather}
t_R\frac{d E_{s}}{dt}=-(\alpha_s + i\Delta \phi_s )E_s + i 2\gamma_s L E_{p1} E_{p2}^* E_{i-} + i 2\gamma_s L E_{p1}^* E_{p2} E_{i+} + i \sqrt{\theta_s P_s}, \label{Eq_CMT_s} \\
t_R\frac{d E_{i+}}{dt}=-(\alpha_{i+} + i\Delta \phi_{i+})E_{i+} + i 2\gamma_{i+} L E_{p1}E_{p2}^* E_s, \label{Eq_CMT_i1} \\
t_R\frac{d E_{i-}}{dt}= -(\alpha_{i-} + i \Delta \phi_{i-})E_{i-} + i 2\gamma_{i-} L E_{p1}^* E_{p2} E_s,
\label{Eq_CMT_i2}
\end{gather}
with their effective detunings given by
\begin{align}
\Delta \phi_s &= (\hat{\omega}_s - \omega_s)t_R - 2\gamma_s L (|E_{p1}|^2 + |E_{p2}|^2), \label{Eq_phis}\\
\Delta \phi_{i+}&=(\hat{\omega}_{i+} - \omega_{i+})t_R - 2\gamma_{i+} L (|E_{p1}|^2 + |E_{p2}|^2), \label{Eq_phii1} \\
\Delta\phi_{i-} &=(\hat{\omega}_{i-} -\omega_{i-})t_R -2\gamma_{i-}L(|E_{p1}|^2 + |E_{p2}|^2) \label{Eq_phii2}.
\end{align}
Because of energy conservation, $\omega_{i\pm}=\omega_s \pm |\omega_{p1} -\omega_{p2}|$. Using Eqs.~\ref{Eq_phip1}, \ref{Eq_phip2}, and \ref{Eq_phis}, we can express the detunings of the two idlers as
\begin{align}
\Delta \phi_{i+}&=(\hat{\omega}_{i+} -\hat{\omega}_{s} - \left |\hat{\omega}_{p1}-\hat{\omega}_{p2} \right |)t_R + \Delta \phi_{s} + (\Delta \phi_{p1} -\Delta \phi_{p2}) - \gamma^{1550}L (|E_{p1}|^2 - |E_{p2}|^2), \label{Eq_phii11} \\
\Delta \phi_{i-}&=(\hat{\omega}_{i-} -\hat{\omega}_{s} + \left |\hat{\omega}_{p1}-\hat{\omega}_{p2} \right |)t_R + \Delta \phi_{s} - (\Delta \phi_{p1} -\Delta \phi_{p2}) + \gamma^{1550}L (|E_{p1}|^2 - |E_{p2}|^2), \label{Eq_phii11}
\end{align}
where we have used the approximation $\gamma_s \approx \gamma_{i+} \approx \gamma_{i-} \approx \gamma^{980}$, since the signal and idlers are close in wavelength. Likewise, for the two pumps, $\gamma_{p1} \approx \gamma_{p2} \approx \gamma^{1550}$. It it worth noting that the frequency detuning $\delta \hat{\omega}_\pm$ defined in the main paper is proportional to the first part of $\Delta \phi_{i\pm}$, as in our experiment (Fig.~3 in the main paper) the two pumps are operated under similar conditions ($P_{p1} \approx P_{p2}$ and $\Delta \phi_{p1} \approx \Delta \phi_{p2}$) and the signal is typically set at the minimum of its transmission ($\Delta \phi_s \approx 0$), thus making the first part the dominant term.

\begin{table}[ht]
\caption{Additional variables used in the coupled mode equations}
\begin{tabular}{|c|c | c|c|}
\hline
Variable name & Description & Equation \\ [0.5ex]
\hline
$\Delta\phi_{p1}$    & Effective detuning for pump 1 & Eq.~4\\
$\Delta\phi_{p2}$    & Effective detuning for pump 2 & Eq.~5\\
$\Delta\phi_{s}$     & Effective detuning for the input signal & Eq.~9\\
$\Delta\phi_{i+}$    & Effective detuning for the blue-shifted idler & Eqs.~10 and 12\\
$\Delta\phi_{i-}$    & Effective detuning for the red-shifted idler & Eqs.~11 and 13\\
$\Omega_{0}$     & Effective nonlinear parameter & $\Omega_{0}=2\gamma^{980}L|E_{p1}E_{p2}|$\\
$\Omega_{1}$     & Asymmetry in idler detuning  & $(\Delta\phi_{i+}-\Delta\phi_{i-})/2$\\
$\Omega_{2}$     & Mean idler detuning  & $(\Delta\phi_{i+}+\Delta\phi_{i-})/2$\\

\hline
\end{tabular}
\end{table}

The coupled mode equations (Eqs.~\ref{Eq_CMT_s}-\ref{Eq_CMT_i2}) generally have three eigenmodes with different eigenfrequencies, readily suggesting mode splitting in the transmission. To gain some insight, we carry out a brief analytical study here. First, we define three parameters: $\Omega_0 \equiv 2\gamma^{980}L|E_{p1}E_{p2}|$, $\Omega_1 \equiv (\Delta \phi_{i+} - \Delta \phi_{i-})/2 $, and $\Omega_2 \equiv (\Delta \phi_{i+} + \Delta \phi_{i-})/2$. Using the result of $\delta \hat{\omega}_\pm$ obtained in the main paper (i.e., $\delta \hat{\omega}_{\pm |\mu|} \approx \pm \delta D_1 |\mu| + \frac{1}{2} \delta D_{2,\mp}\mu^2$, where $\mu$ stands for the azimuthal order difference between the two pumps), we arrive at
\begin{gather}
\Omega_1  \approx  \delta D_1 |\mu| t_R  -\frac{1}{2}D_2^{1550} \mu^2 t_R + \Delta \phi_{p1} -\Delta \phi_{p2} - \gamma^{1550}L (|E_{p1}|^2 - |E_{p2}|^2),  \label{Eq_Omega1}\\
\Omega_2  \approx \frac{1}{2}D_2^{980} \mu^2 t_R. \label{Eq_Omega2}
\end{gather}
One can easily identify the factors affecting $\Omega_1$ are: the FSR mismatch between the $1550$ nm and $980$ nm bands, the second order dispersion in the $1550$ nm band, and the experimental operating conditions of the two $1550$ nm pumps (especially the symmetry). $\Omega_2$, on the other hand, is generally small and purely determined by the dispersion of the $980$ nm band (in particular, it is independent of the two pump settings). Here we approximate $\Omega_2 \approx 0$, and the three eigenfrequencies can be calculated as
\begin{align}
\hat{\omega}_{e1} &\approx \hat{\omega}_s,\\
\hat{\omega}_{e2} &\approx \hat{\omega}_s + \frac{1}{t_R}\sqrt{2\Omega_0^2 + \Omega_1^2},\\
\hat{\omega}_{e3} &\approx \hat{\omega}_s - \frac{1}{t_R}\sqrt{2\Omega_0^2 + \Omega_1^2}.
\end{align}
It is also interesting to find the corresponding eigenmodes, which are
\begin{align}
\ket{\mathbf{e}_1} &= N_1\left(\frac{1}{\sqrt{1+2\eta^2}} \ket{\mathbf{s}}-  \frac{\eta}{\sqrt{1+2\eta^2}} \ket{\mathbf{i}+} + \frac{\eta}{\sqrt{1+2\eta^2}} \ket{\mathbf{i}-} \right),  \label{Eq_e1} \\
\ket{\mathbf{e}_2}&= N_2 \left (\ket{\mathbf{s}} + \frac{\eta}{\sqrt{1+2\eta^2}-1} \ket{\mathbf{i}+}
+ \frac{\eta}{\sqrt{ 1+ 2 \eta^2}+1} \ket{\mathbf{i}-} \right ), \label{Eq_e2} \\
\ket{\mathbf{e}_3}&= N_3 \left (\ket{\mathbf{s}} - \frac{\eta}{\sqrt{1+2\eta^2}+1} \ket{\mathbf{i}+}
-\frac{\eta}{\sqrt{ 1+ 2 \eta^2}-1} \ket{\mathbf{i}-} \right ), \label{Eq_e3}
\end{align}
where $\eta \equiv \Omega_0/\Omega_1$ and $N_k$ is a normalization factor for $\ket{\mathbf{e}_k}$ to ensure $\left \langle{\mathbf{e}_k}|{\mathbf{e}_k}\right\rangle = 1$ ($k=1,2,3$). Generally, the state of photons travelling inside the cavity will be a superposition of the these three eigenmodes, with the coefficients determined by  parameters including $\alpha_s$ (cavity linewidth) and [$\Omega_0$ $\Omega_1$  $\Omega_2$]. Down to single photon level, the square of the coefficient associated with each eigenmode can be interpreted as the probability of finding the injected photon in this particular mode.

\begin{center}
\begin{figure*}
\begin{center}
\includegraphics[width=0.85\linewidth]{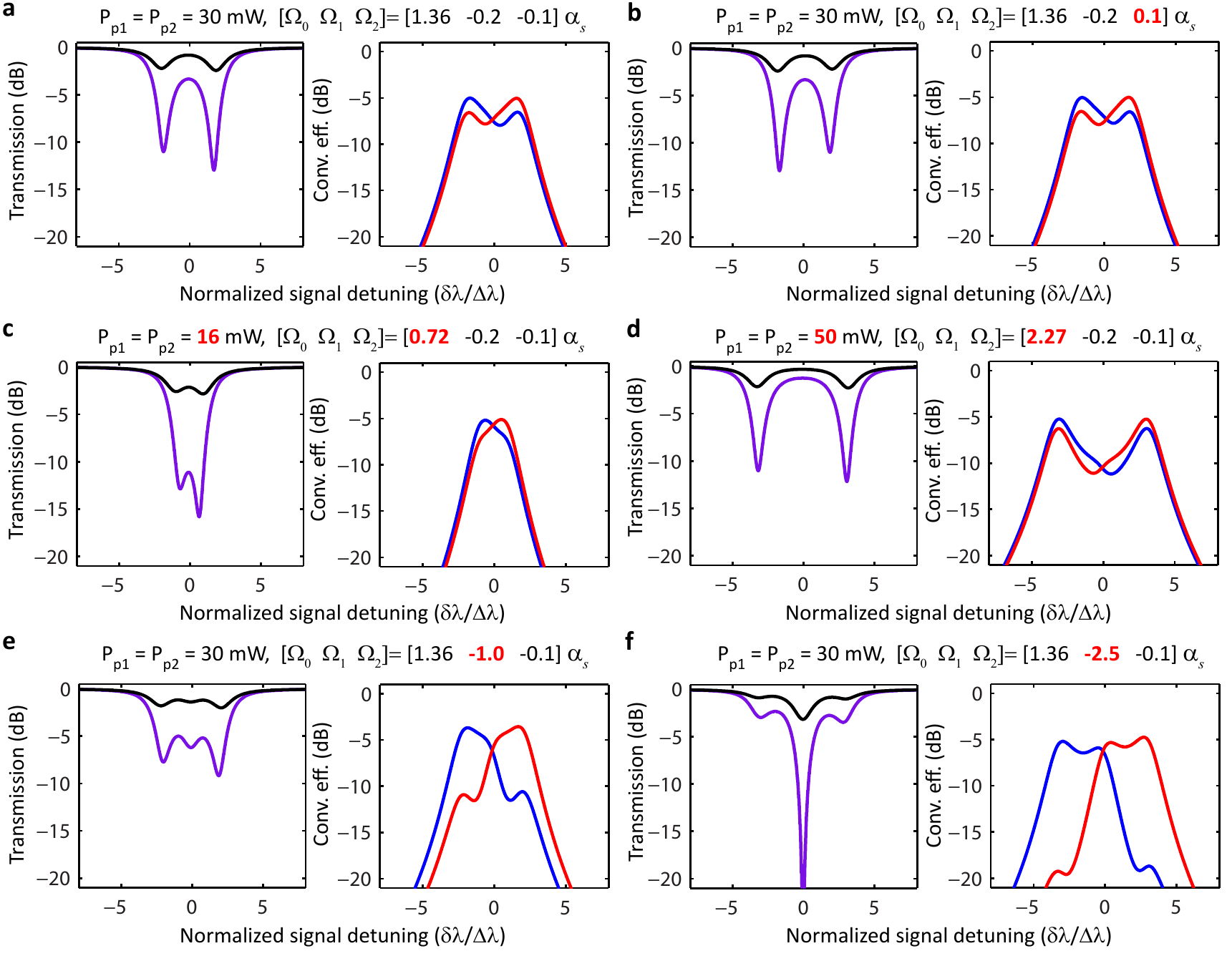}
\caption{\textbf{Coupled mode simulations for the 980 nm intraband conversion}. \textbf{a} Reference example of the parameter variations of [$\Omega_0 \ \Omega_1 \ \Omega_2$] in units of $\alpha_s$ (cavity loss rate at the signal resonance). The varied parameter in other cases is highlighted. For each case, the left figure shows the transmission (purple: signal only; black: signal plus two idlers) and the right figure shows the on-chip conversion efficiency (blue for $i+$ and red for $i-$) as a function of normalized signal detunings (-$\Delta \phi_s /\alpha_s = \delta \lambda/\Delta \lambda$ where $2 \Delta \lambda \approx 4$ pm is the linewidth of the signal resonance). \textbf{b}, Changing the sign of $\Omega_2$ will reverse the symmetry between the two split modes in the transmission scan. \textbf{c}-\textbf{d},Varying $\Omega_0$ (i.e., the pump power): mode splitting increases with the pump power. \textbf{e}-\textbf{f}, Varying $\Omega_1$: nonzero $|\Omega_1|$ removes the degeneracy in the conversion efficiency between $i+$ and $i-$; especially, a relatively large $|\Omega_1|$ can make one idler much stronger than the other at its optimum signal detuning.}
\label{Fig_CMT}
\end{center}
\end{figure*}
\end{center}

Next, numerical simulations are carried out in Fig.~\ref{Fig_CMT} to investigate the effect of different combinations of  [$\Omega_0\  \Omega_1\ \Omega_2$] to the frequency conversion process (the two pumps are assumed to have equal power and detuning). Figure \ref{Fig_CMT}a plots the reference example, which corresponds to the device under study for a pump power of $30$ mW per pump. The transmission scan of the signal resonance (left figure) and the conversion efficiency for the blue and red idlers (right figure) as a function of signal detuning are similar to the experimental results (Fig.~3b, case IV) and the LLE simulation (Fig.~4c) shown in the main paper. The asymmetric mode splitting observed in the transmission scan is due to a nonzero $\Omega_2$, which is negative for our device (see Eq.~\ref{Eq_Omega2} and note $D_2^{980}<0$ since the resonator shows normal dispersion in both $1550$ nm and $980$ nm bands). If $\Omega_2$ is positive, as illustrated by Fig.~\ref{Fig_CMT}b, the asymmetry in the transmission scan is mirrored in the spectrum, whereas a symmetric mode splitting is expected if $\Omega_2$ is zero.

Figures \ref{Fig_CMT}c-d explore how the pump power affects the frequency conversion process (varying $\Omega_0$), which shows that the maximum conversion efficiency for both blue and red idlers is limited to $\approx$~$-5$ dB for pump powers larger than $16$ mW. In addition, as the pump power increases, the mode splitting becomes stronger and the optimum signal detuning corresponding to the maximum conversion efficiency for the blue and red idlers shifts away from origin. Therefore, if we choose to use a fixed signal detuning (such as zero), the conversion efficiency will start to decrease with the pump power after reaching its optimum value~\cite{ref:Kumar_FWBS_theory_OL_SM}. Finally, Figs.~\ref{Fig_CMT}e-f study the effect of $\Omega_1$, that is, the asymmetry in the idler detunings with respect to the nearest cavity modes. According to Eq.~\ref{Eq_Omega1}, its absolute value can be easily comparable to the cavity loss rate if the separation between the two $1550$ nm pumps ($|\mu|$) or the FSR mismatch between the $1550$ nm and $980$ nm bands is large enough. In addition, its value can change significantly if the two pumps are operated under different conditions (such as $P_{p2}\gg P_{p1}$ as in the wideband conversion experiment shown in the main paper). The simulation shows that as $|\Omega_1|$ increases (Fig.~\ref{Fig_CMT}e), we start to see three dips in the transmission scan, corresponding to the three eigenmodes of the coupled system (a similar spectrum has been observed in the experiment for a different device, though the data not is shown here). Moreover, unlike the previous examples (Figs.~\ref{Fig_CMT}a-d) where the blue and red idlers show similar conversion efficiencies, in this case only one idler can reach its maximum conversion efficiency while the other one is much weaker. Further increasing $|\Omega_1|$ (Fig.~\ref{Fig_CMT}f) results in a bigger contrast between the blue and red idlers if one of them is set at its optimum conversion efficiency. However, if $|\Omega_1|$ is too large ($> 5 \alpha_s$ for the pump power of $30$ mW), the maximum achievable conversion efficiency for both idlers drops significantly ($<10~\%$), and the transmission scan of the signal resonance starts to look like the linear case.

\subsection{Coupled mode equations for wideband frequency conversion}
The coupled mode equations for the wideband frequency conversion can be derived following a similar fashion as in the previous subsection. Especially, as demonstrated by Fig.~5d in the main paper, the pump mixing is absent in the wideband conversion case since the two pumps are widely separated. Consequently, other than the two pumps,  there are typically four modes: the signal (s), the auxiliary tone generated from the mixing of the signal and its nearby pump (s'), and two idlers ($i+$ and $i-$). The coupled mode equations developed here will only consider these four modes, which means we neglect higher order idlers generated from the mixing of $i\pm$ with its nearby pump.

\begin{center}
\begin{figure*}
\begin{center}
\includegraphics[width=0.36\linewidth]{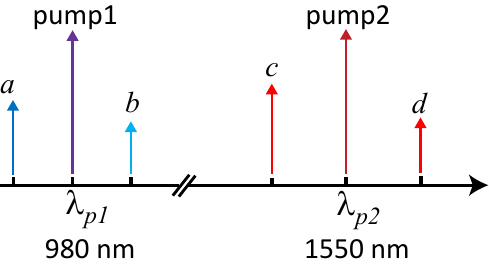}
\caption{\textbf{Schematic of the wideband conversion process.} This schematic shows the four modes under consideration in the coupled mode equations for the wideband  frequency conversion. For the upconversion, $c$ is the signal, $d$ is the auxiliary tone resulting from the wave mixing between $c$ and pump 2, and $a$ and $b$ are $i+$ and $i-$, respectively. For the downconversion, $a$ is the signal, $b$ is the auxiliary tone resulting from the wave mixing between $a$ and pump 1, and $c$ and $d$ are $i+$ and $i-$, respectively.}
\label{Fig_Wideband}
\end{center}
\end{figure*}
\end{center}

To treat both the upconversion and downconversion, we label the two modes in the $980$ nm band as $a$ and $b$, and the two modes in the $1550$ nm band as $c$ and $d$ (see Fig.~\ref{Fig_Wideband}). For the upconversion case, [$a$ $b$ $c$ $d$] correspond to [$i+$ $i-$ $s$ $s'$], respectively. For the downconversion case, [$a$ $b$ $c$ $d$] correspond to [$s$ $s'$ $i+$ $i-$], respectively.

First, we consider the case without the signal input field driving its cavity resonance. The coupled mode equations for these four modes can be written as
\begin{align}
t_R \frac{dE_a}{dt}&=-(\alpha_{p1} + i\Delta \phi_a)E_a + i\gamma_{p1}L E_{p1}^2E_b^* + i2\gamma_{p1}LE_{p1}(E_{p2}E_d^* + E_{p2}^*E_c),\label{Eq_Wa} \\
t_R \frac{dE_b}{dt}&=-(\alpha_{p1} + i\Delta \phi_b)E_b + i\gamma_{p1}L E_{p1}^2E_a^* + i2\gamma_{p1}LE_{p1}(E_{p2}E_c^* + E_{p2}^*E_d),\label{Eq_Wb} \\
t_R \frac{dE_c}{dt}&=-(\alpha_{p2} + i\Delta \phi_c)E_c + i\gamma_{p2}L E_{p2}^2E_d^* + i2\gamma_{p2}LE_{p2}(E_{p1}E_b^* + E_{p1}^*E_a),\label{Eq_Wc} \\
t_R \frac{dE_d}{dt}&=-(\alpha_{p2} + i\Delta \phi_d)E_d + i\gamma_{p2}L E_{p2}^2E_c^* + i2\gamma_{p2}LE_{p2}(E_{p1}E_a^* + E_{p1}^*E_b),\label{Eq_Wd}
\end{align}
where for simplicity we have approximated $\alpha$ and $\gamma$ of each wave as the values of its nearby pump (e.g., $\gamma_{a,b}=\gamma_{p1}$ and $\gamma_{c,d}=\gamma_{p2}$). In addition, the effective detunings of these modes are obtained as
\begin{align}
\Delta \phi_a &= (\hat{\omega}_a -\omega_a)t_R - 2\gamma_{p1}L ( |E_{p1}|^2 + |E_{p2}|^2),\label{Eq_phia}\\
\Delta \phi_b &= (\hat{\omega}_b -\omega_b)t_R -2 \gamma_{p1}L (|E_{p1}|^2 + |E_{p2}|^2),\label{Eq_phib}\\
\Delta \phi_c &= (\hat{\omega}_c -\omega_c) t_R -2 \gamma_{p2} L(|E_{p1}|^2 + |E_{p2}|^2),\label{Eq_phic}\\
\Delta \phi_d &=(\hat{\omega}_d -\omega_d)t_R - 2\gamma_{p2}L(|E_{p1}|^2 + |E_{p2}|^2).\label{Eq_phid}
\end{align}
Similar to the 980~nm intraband conversion, the energy conservation condition (see Fig.~5 in the main paper) is used to express the detunings of the idlers as a function of the signal and pump detunings, which are free parameters to be varied in the experiment. For example, for the upconversion case, energy conservation requires $\omega_a=\omega_{p1} +|\omega_c -\omega_{p2}|$, $\omega_b=\omega_{p1}-|\omega_c -\omega_{p2}|$, and $\omega_d=2\omega_{p2} -\omega_c$. With some algebra, we arrive at
\begin{gather}
\Delta \phi_a= [(\hat{\omega}_a -\hat{\omega}_{p1}) -(\hat{\omega}_c-\hat{\omega}_{p2})]t_R + \Delta \phi_c +(\Delta \phi_{p1} -\Delta \phi_{p2})-(\gamma_{p1}L|E_{p1}|^2 - \gamma_{p2}L|E_{p2}|^2), \label{Eq_upphia2}\\
\Delta \phi_b=[(\hat{\omega}_b -\hat{\omega}_{p1}) +(\hat{\omega}_c-\hat{\omega}_{p2})]t_R - \Delta \phi_c +(\Delta \phi_{p1} + \Delta \phi_{p2}) - (\gamma_{p1}L|E_{p1}|^2 + \gamma_{p2}L|E_{p2}|^2), \label{Eq_upphib2}\\
\Delta \phi_d = [\hat{\omega}_d + \hat{\omega}_c -2\hat{\omega}_{p2}]t_R - \Delta \phi_c + 2\Delta \phi_{p2} -2\gamma_{p2}L|E_{p2}|^2.  \label{Eq_upphic2}
\end{gather}
In the same way, for the downconversion case ($a$ is the signal), we have $\omega_c = \omega_{p2} + |\omega_a -\omega_{p1}|$, $\omega_d = \omega_{p2} - |\omega_a - \omega_{p1}|$, and $\omega_b= 2\omega_{p1} - \omega_a$. The effective detunings of these modes can then be derived as
\begin{gather}
\Delta \phi_c= [(\hat{\omega}_c -\hat{\omega}_{p2}) -(\hat{\omega}_a-\hat{\omega}_{p1})]t_R + \Delta \phi_a -(\Delta \phi_{p1} -\Delta \phi_{p2})+ (\gamma_{p1}L|E_{p1}|^2 - \gamma_{p2}L|E_{p2}|^2), \label{Eq_downphic}\\
\Delta \phi_d=[(\hat{\omega}_d -\hat{\omega}_{p2}) +(\hat{\omega}_a-\hat{\omega}_{p1})]t_R - \Delta \phi_a +(\Delta \phi_{p1} + \Delta \phi_{p2}) - (\gamma_{p1}L|E_{p1}|^2 + \gamma_{p2}L|E_{p2}|^2), \label{Eq_downphid}\\
\Delta \phi_b = [\hat{\omega}_a + \hat{\omega}_b -2\hat{\omega}_{p1}]t_R - \Delta \phi_a + 2\Delta \phi_{p1} -2\gamma_{p1}L|E_{p1}|^2.  \label{Eq_downphib}
\end{gather}

Now we add back the driving field for the signal. By defining a vector $\tilde{E}_r= [E_a\ E_b^* \ E_c\ E_d^*]^T$ (the subscript $r$ is to denote that they are fields inside resonator and the superscript $T$ stands for transpose of a matrix), we can write Eqs.~\ref{Eq_Wa}-\ref{Eq_Wd} into a more compact form as
\begin{equation}
t_R \frac{d\tilde{E}_r}{dt} = \tilde{M}\tilde{E}_r + i\sqrt{\tilde{\Theta}}\tilde{E}_{\text{in}}, \label{Eq_dMt}
\end{equation}
where the matrix $\tilde{M}$ is defined by
\begin{equation}
\tilde{M}\equiv \begin{bmatrix}  -\alpha_{p1} - i\Delta \phi_{a} &  i \gamma_{p1}LE_{p1}^2 & i2\gamma_{p1}L E_{p1} E_{p2}^* & i2\gamma_{p1}L E_{p1}E_{p2} \\
-i\gamma_{p1}LE_{p1}^{*2} & -\alpha_{p1} + i\Delta \phi_b & -i2\gamma_{p1}LE_{p1}^* E_{p2}^*& -i2\gamma_{p1}LE_{p1}^*E_{p2} \\ i2\gamma_{p2}LE_{p1}^*E_{p2}& i2\gamma_{p2}LE_{p1}E_{p2} & -\alpha_{p2} - i \Delta \phi_c & i\gamma_{p2}LE_{p2}^2 \\ -i2\gamma_{p2}LE_{p1}^*E_{p2}^* & -i2\gamma_{p2}LE_{p1}E_{p2}^*& -i\gamma_{p2}LE_{p2}^{*2}& -\alpha_{p2} + i\Delta \phi_d \end{bmatrix}, \label{Eq_MW}
\end{equation}
$\tilde{\Theta}$ is a diagonal matrix defined as $\tilde{\Theta}\equiv  \text{diag} (\theta_{p1}\ \theta_{p1}\ \theta_{p2}\ \theta_{p2})$, and $\tilde{E}_{\text{in}}$ is the driving field. For the upconversion case (mode $c$ is the signal), $\tilde{E}_{\text{in}}= [0\ 0\ \sqrt{P_s}\ 0]^T$ and for the downconversion case (mode $a$ is the signal), $\tilde{E}_{\text{in}}= [\sqrt{P_s}\ 0\ 0\ 0]^T$. The steady state solution can be found from Eq.~\ref{Eq_dMt} as $\tilde{E}_r= - i\tilde{M} ^{-1} \sqrt{\tilde{\Theta}} \tilde{E}_{\text{in}}$ for the fields inside cavity and $\tilde{E}_{\text{WG}} = \tilde{E}_{\text{in}} + i\sqrt{\tilde{\Theta}} \tilde{E}_r$ for the fields at the output of the waveguide. Note that each component of $|\tilde{E}_{\text{WG}}|^2$ represents the power of the corresponding mode, which has to be converted to photon flux for the calculation of the the on-chip conversion efficiency. Simulation examples are provided in Fig.~\ref{Fig_Upconv}c for the upconversion case and Fig.~6b in the main paper for the downconversion case, both showing a good agreement with the experimental results.

\subsection{Modified LLE method for FWM-BS}
In this subsection, we will discuss a way to adapt the conventional LLE method for the FWM-BS process. So far, the coupled mode equations presented only consider a few modes. For example, for the intraband conversion in the $980$ nm band, we have neglected the pump mixing in the $1550$ nm band as well as  high-order idler generation in the $980$ nm band. It is still possible to include these modes in the coupled mode equations, although the process is typically very manual. By contrast, the LLE formalism can automatically include all resonant modes for the entire wavelength band of interest, making it a convenient tool to study the interaction of multiple modes.

As pointed out at the beginning of this section, in the standard LLE formalism we only set the detuning of the pump ($\delta_0$ in Eq.~\ref{Eq_LLE}) which is the only driving field~\cite{ref:Coen_LLE_SM, ref:Chembo_coupled_mode_combs_SM}. In addition, the resonance frequencies of all the modes are referenced to an equally spaced frequency grid $\{\omega_0 + D_1 \mu\}$, with $D_1$ corresponding to the FSR of the pump resonance. In the FWM-BS process, since we have three driving fields (two pumps and one signal), it is not straightforward to allow these three waves have their own detunings independent from the others. In addition, it is also not trivial to choose $D_1$. For example, even if we decide $D_1$ should be the FSR of the resonator, we still face the choice of the FSR in the $1550$ nm or the $980$ nm band.

The solution we find is to implement the LLE formalism based on Eq.~\ref{Eq_CMT}, whose detuning parameter ($\delta_m = \hat{\omega}_m -\omega_m$) is allowed to vary across different resonant modes. Thus, we can set $\delta_m$ corresponding to the two pumps and the signal. The detunings of the remaining modes are determined from the following procedure. First, we obtain resonance frequencies ($\{\hat{\omega}_m\}$) of all the modes from numerical simulations based on an eigenfrequency mode solver for the microresonator under study; second, the equally spaced frequency grid ($\{\omega_m\}$) is set up in the $1550$ nm and $980$ nm bands separately before joining together. To make it clear, we use the $980$ nm intraband conversion as an example (Fig.~S4). Assuming the $1550$ nm pumps are accommodated by two resonances adjacent to each other (i.e., $m_{p1}-m_{p2}=1$), the frequency grid in the $980$ nm band is chosen as $\omega_m =\omega_s + (m-m_s) |\omega_{p1} - \omega_{p2}|$, where $m_s$ is the azimuthal order of the signal resonance. By doing so, the signal and all the generated idlers automatically fall on the frequency grid (energy conservation satisfied). Similar to the situation in the coupled mode equations, here $\omega_s$, $\omega_{p1}$ and $\omega_{p2}$ can be replaced by their respective detunings ($\omega_m = \hat{\omega}_m - \delta_m$). Likewise, in the $1550$ nm band, we can start with either pump and expand the grid with the same spacing (i.e., $|\omega_{p1} - \omega_{p2}|$). This allows the two $1550$ nm pumps and secondary pumps generated from pump mixing to be on the grid, since the energy conservation is automatically satisfied. Finally, we expand the frequency grid in the $980$ nm band and the one in the $1550$ nm band until they start to overlap the middle of the total frequency span (typically we consider a wavelength range from $900$ nm to $1700$ nm, so the midpoint is around $1200$ nm). Admittedly, there could be some error for the detuning of the joint point (basically $\omega_m$ extended from the $980$ nm grid is not necessarily equal to that from the $1550$ nm grid), but generally this mode is frequency mismatched and far away from the wavelength bands that interest us. As a result, we end up with a frequency grid that spans the whole spectral range, and all the modes of interest are on the grid. Following the same reasoning, if the two $1550$ nm pumps are accommodated by resonances with their azimuthal order difference $N=|m_{p1}- m_{p2}|>1$, we should take the grid spacing as $|\omega_{p1} - \omega_{p2}|/N$. Similarly, for the wideband conversion case, the grid spacing is taken as the frequency difference between the signal and its nearby pump, divided by the difference of their azimuthal orders. Therefore, we can clearly see from this process that the frequency grid spacing is not exactly the FSR of the resonator, with the deviation contributed by nonzero detuning and high-order dispersions.

From the discussion, it should be clear by now that the modified LLE method described above is essentially a systematic way to implement the coupled mode equations discussed in previous subsections (A and B), so that all the resonant modes can be included easily. Other than the detuning, parameters such as $\alpha$, $\gamma$, and $\theta$ in Eq.~\ref{Eq_CMT} can (and should) be frequency dependent. To solve Eq.~\ref{Eq_CMT} numerically, we use the well-known split-step Fourier method, i.e., we calculate the frequency-dependent terms (Eq.~\ref{Eq_CMT} without the nonlinear Fourier transformed term) in the frequency domain and then compute the nonlinear term in the time domain. The implementation is similar to that described in Ref.~\onlinecite{ref:Agrawal_NFO}.

\begin{center}
\begin{figure*}
\begin{center}
\includegraphics[width=0.8\linewidth]{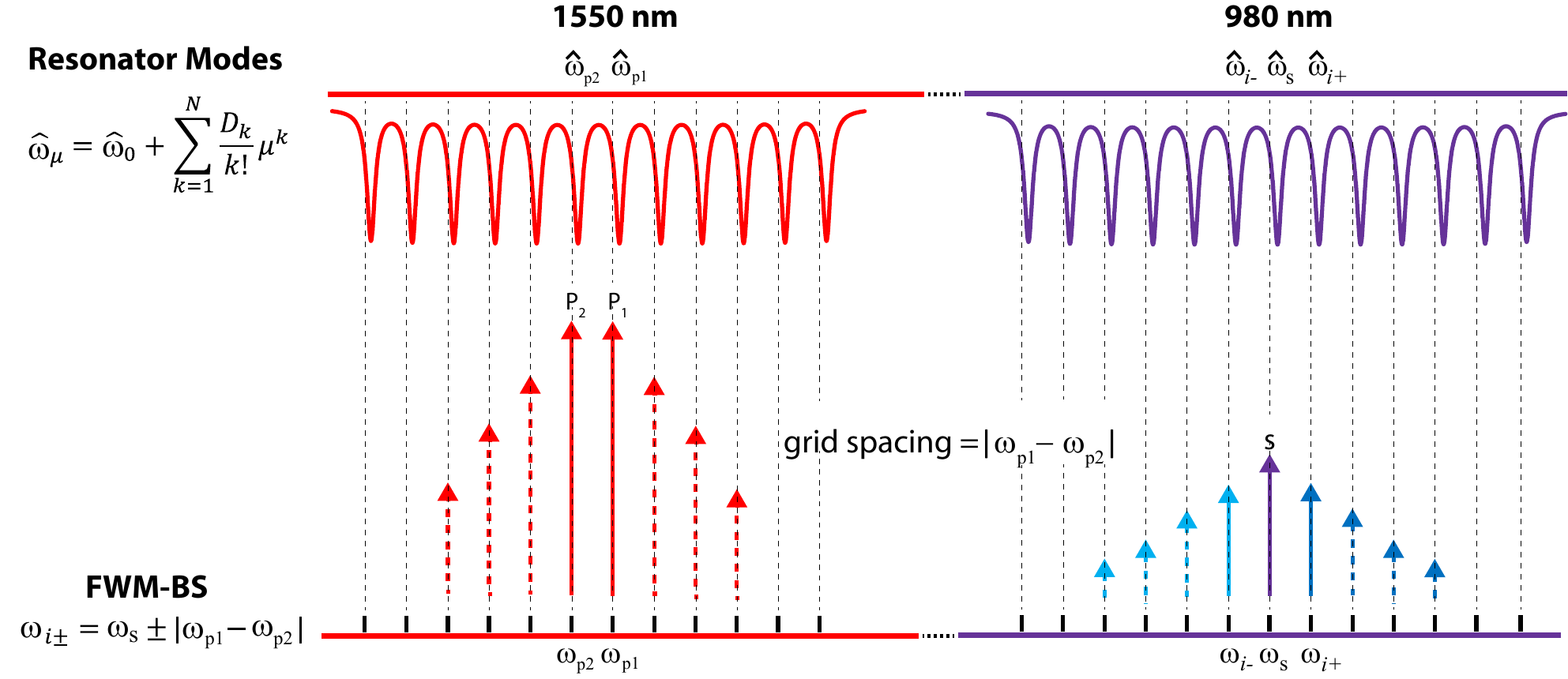}
\caption{\textbf{Schematic of the frequency grid used in the LLE simulation.} The frequency grid shown here corresponds to the $980$ nm intraband conversion with two 1550~nm band pumps accommodated by two adjacent resonances. The vertical dashed arrows in the $1550$ nm and $980$ nm band represent the secondary pumps generated from the pump mixing and high-order idlers, respectively. The resonator modes are illustrated above, indicating the resonance frequencies are not exactly on the grid due to non-zero detuning and high-order dispersions.}
\label{Fig_LLE}
\end{center}
\end{figure*}
\end{center}

\subsection{Calculation of Kerr nonlinear coefficient}
The purpose of this subsection is to provide a brief discussion on the calculation of the Kerr nonlinear coefficient $\gamma$ in a nanophotonic structure, as we have noticed there are several different versions existing in the literature ~\cite{ref:Agrawal_NFO_SM,ref:Lin_Painter_gamma_OE_SM}. We will consider only one wave (the pump) here, as it is fairly straightforward to generalize the result to multiple waves ~\cite{ref:Lin_photon_pair_OE_SM,ref:Popovic_Nonlinear_OE_SM}. The most common version for $\gamma$ is probably $\gamma= \frac{n_2\omega}{cA_{\text{eff}}}$ (Ref.~\onlinecite{ref:Agrawal_NFO}), where $n_2$ is the Kerr nonlinear refractive index ($n_2 \approx 2.5 \times 10^{-19}\ \text{m}^2 \text{W}^{-1}$ for Si$_3$N$_4$), and $A_{\text{eff}}$ is the effective mode area given by
\begin{equation}
A_{\text{eff}} = \frac{\left(\int\int |\mathbf E(x,y)|^2\ dx dy \right )^2}{\int\int_{\text{core}}|\mathbf E(x,y)|^4\ dxdy}, \label{Eq_Aeff}
\end{equation}
where $\mathbf E(x,y)$ is the electric field of the mode under consideration, and the subscript in the integral denotes the integral over the nonlinear (core) material. This result is typically applied to fibers or waveguides with low index contrast materials.

It is expected that Eq.~\ref{Eq_Aeff} requires some modification when applied to nanophotonic structures with high index contrast material systems (such as Si$_3$N$_4$/SiO$_2$, Si/SiO$_2$ or AlGaAs/SiO$_2$). For the  $\chi^{(3)}$ process in a microresonator, it has been shown that the effective mode volume is given by ~\cite{ref:Lin_Painter_gamma_OE_SM}
\begin{equation}
V_{\text{eff}}=\frac{\left(\int\int\int \epsilon_r(\mathbf{r}) |\mathbf E(\mathbf{r})|^2\ d^3\mathbf{r}\right)^2}{\int\int\int_{\text{core}}\epsilon_r^2(\mathbf{r}) |\mathbf E(\mathbf{r})|^4\ d^3\mathbf{r}},
\label{Eq_Veff}
\end{equation}
where $\epsilon_r$ is the relative permittivity. Following this, one may naturally expect that for a waveguide,
\begin{equation}
\bar{A}_{\text{eff}}=\frac{\left(\int\int \epsilon_r(x,y) |\mathbf E(x,y)|^2\ dxdy \right)^2}{\int\int_{\text{core}}\epsilon_r^2(x,y) |\mathbf E(x,y)|^4\ dxdy}.
\label{Eq_Beff}
\end{equation}

On the other hand, we can start with the variational principle \cite{ref:Haus_book_SM}, that is,
\begin{equation}
\delta \beta = \frac{\omega \int \int_{\text{core}} \mathbf{E}^*(x,y)\cdot \mathbf{P}(x,y)\ dxdy}{4P_o},
\label{Eq_vp}
\end{equation}
where $\mathbf P(x,y)$ is the dielectric perturbation and $P_o$ is the power of the waveguide. For the $\chi^{(3)}$ process involving only one wave, we have $\mathbf P(x,y)= 3\epsilon_0 \chi^{(3)}_{1111} |\mathbf E(x,y)|^2 \mathbf E(x,y)/4$, where $\epsilon_0$ is the vacuum permittivity and $\chi^{(3)}_{1111}$ is related to $n_2$ as $n_2= 3 \chi^{(3)}_{1111}/(4 n_0^2 \epsilon_0 c)$ (Ref.~\onlinecite{ref:Lin_Painter_Agrawal}), with $n_0$ being the refractive index of the material. One can easily observe that the nonlinear perturbation results in a power-dependent $\delta\beta$, which is recognized as the self phase modulation. Hence, $\delta \beta = \gamma P_o$. As a result,
we end up with $\gamma=\frac{n_2 \omega}{c\tilde{A}_{\text{eff}}}$, with a new mode area $\tilde{A}_{\text{eff}}$ defined as
\begin{equation}
\tilde{A}_{\text{eff}}= \frac{4 P_o^2}{n_0^2 \epsilon_0^2 c^2 \int \int_{\text{core}} |\mathbf E(x,y)|^4\ dxdy}. \label{Eq_Qeff}
\end{equation}
Since the power of the waveguide mode is given by an integral of mode energy density over the waveguide cross section multiplied by the group velocity, we can simplify $\tilde{A}_{\text{eff}}$ as
\begin{equation}
\tilde{A}_{\text{eff}}=\left(\frac{n_0}{n_g}\right )^2 \frac{\left(\int\int \epsilon_r(x,y) |\mathbf E(x,y)|^2\ dxdy \right)^2}{\int\int_{\text{core}}\epsilon_r^2(x,y) |\mathbf E(x,y)|^4\ dxdy},
\label{Eq_Qeff2}
\end{equation}
which shows there is an additional factor $(n_0/n_g)^2$ compared to $\bar{A}_{\text{eff}}$ in Eq.~\ref{Eq_Beff}.

We want to point out Eq.~\ref{Eq_Veff} is derived based on the same variational principle as Eq.~\ref{Eq_Qeff2}. On the other hand, we believe it is not strictly valid to go from Eq.~\ref{Eq_Veff} to Eq.~\ref{Eq_Beff} for two reasons. First, Eq.~\ref{Eq_Veff} is derived for modes with normalized energy, while for waveguides we are dealing with modes normalized by power. The conversion from energy to power introduces a factor proportional to the round-trip time ($t_R=Ln_g/c$) for the $\chi^{(3)}$ nonlinear term. Second, what has been calculated for the waveguide case is $\delta \beta$ caused by the nonlinear perturbation, while the counterpart for the cavity case is the frequency shift $\delta \omega$. They are related by $\delta \beta \approx n_g \delta \omega/c$. Thus, it seems the $(n_0/n_g)^2$ factor can be explained. Nevertheless, for the Si$_3$N$_4$ waveguides (that form the ring resonator) studied in this work, the numerical value of $\bar A_{\text{eff}}$ is almost the same as $\tilde A_{\text{eff}}$, since $n_g \approx n_0 \approx 2$. In comparison, the calculated $\gamma$ is about $35~\%$ higher than the result based on Eq.~\ref{Eq_Aeff}, suggesting it is important to choose the right mode area for the numerical simulation.

\section{Parameters used in the simulation}
We list the parameters used in the simulations in the following table.
\begin{table}[h!]
\caption{Parameters used in the simulations}
\begin{tabular}{|c|c | c|c|}
\hline
Parameter&1550 nm band&980 nm band& Unit\\ [0.5ex]
\hline
\noalign{\vskip 0.3mm}
$Q_i$ & $4.5\times 10^5$& $9.0 \times 10^5$ & -\\
$Q_c$ & $2.3\times 10^5$& $3.3 \times 10^5$ & -\\
$Q_L$ & $1.5\times 10^5$& $2.4 \times 10^5$ & -\\
$D_1/2\pi$ & $572.39$ & $572.25 $ &GHz \\
$D_2/2\pi$ & $-31.04$ & $-3.43 $  &MHz \\
$D_3/2\pi$ & $0.71$ &$-0.30$ & MHz \\
$\gamma$ & $2.07$ & $3.32$ & W$^{-1}$m$^{-1}$\\
\hline
\end{tabular}
\end{table}

\section{Transmission measurements}
\begin{center}
\begin{figure*}
\begin{center}
\includegraphics[width=0.85\linewidth]{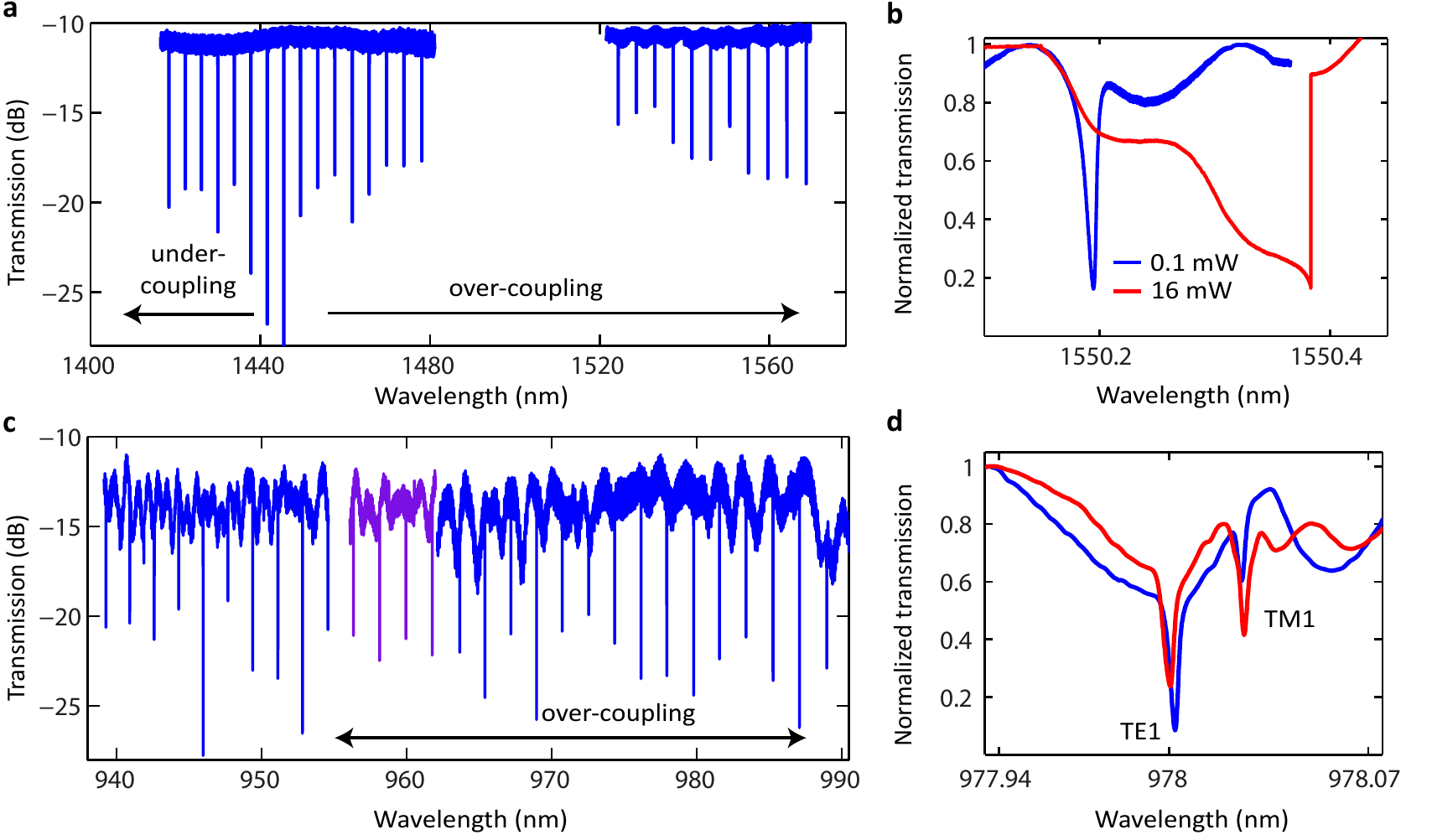}
\caption{\textbf{Transmission measurement of the microring device}. \textbf{a}, Transmission of the microring measured with tunable lasers covering wavelength ranges from $1415$ nm to $1480$ nm and $1520$ nm to $1570$ nm.  \textbf{b}, Normalized transmission scan of the resonance around $1550$ nm at a low power ($0.1$ mW, showing a very slight thermal triangle) and high power ($16$ mW, showing a very strong thermal triangle). \textbf{c}, Transmission of the microring measured with tunable lasers (blue lines) from $939$ nm to $955$ nm and $962$ nm to $990$ nm. The resonances in the wavelength gap (purple line) are measured by a third tunable laser covering wavelength range from $915$ nm to $985$ nm. \textbf{d}, Normalized transmission scan of the $978$ nm resonance (power $= 0.1$ mW) when the polarization is TE (blue) and slightly away from TE (red).}
\label{Fig_Trans}
\end{center}
\end{figure*}
\end{center}

To characterize the resonances of the microring device in the $1550$ nm band, two tunable lasers covering  wavelength ranges from $1415$ nm to $1480$ nm and $1520$ nm to $1570$ nm are employed. Figure \ref{Fig_Trans}a shows the measured transmission (defined as the power collected by the lens fiber from the waveguide output normalized by the power sent to the lens fiber that couples to the waveguide input). As can be seen, even though the microring is multimode, only the fundamental TE mode is observed in the transmission spectrum, thanks to the pulley coupling design~\cite{ref:Adibi_grp_pulley_SM}. Figure \ref{Fig_Trans}b shows the transmission scan of the resonance around $1550$ nm. At low powers, the resonance shape is close to Lorentzian; as the power increases, thermo-optic dispersion quickly results in the swept-wavelength lineshape following the commonly observed thermal triangle.

Figure \ref{Fig_Trans}c shows the transmission of the $980$ nm band, whose characterization involves three tunable lasers. The first two lasers, which have been used as the signal in the downconversion experiment (Fig.~6a in the main paper), cover wavelength ranges from $930$ nm to $955$ nm and $960$ nm to $990$ nm. The third laser, which has been used as the $980$ nm pump for the wideband conversion experiment, has a tunable wavelength range from $915$ nm to $985$ nm and a relatively large output power ($\approx 45$ mW). Considering $\approx 6.5$ dB coupling loss at the input facet, the maximum on-chip pump power is limited to $<10$ mW. Similar to the $1550$ nm spectrum, the transmission of the $980$ nm band shows that only the fundamental TE mode is well-coupled while other modes are suppressed. However, as illustrated by Fig.~\ref{Fig_Trans}d, the fundamental TE mode can interact with some other mode at certain wavelengths, which shifts its resonance frequency as shown in Fig.~2f in the main paper. By changing the polarization state of the input light, we find we can increase the extinction ratio of the mode that couples to the TE resonance, suggesting this mode has a different polarization. Through a careful phase-matching analysis, we believe the mode is most likely to be the fundamental TM mode. Such excitation is possible because the TE waveguide mode has a nonzero field overlap with the TM mode of the resonator, though generally the coupling is weak and only strong at certain wavelengths due to an accidental phase matching.

\begin{center}
\begin{figure*}[h!]
\begin{center}
\includegraphics[width=0.85\linewidth]{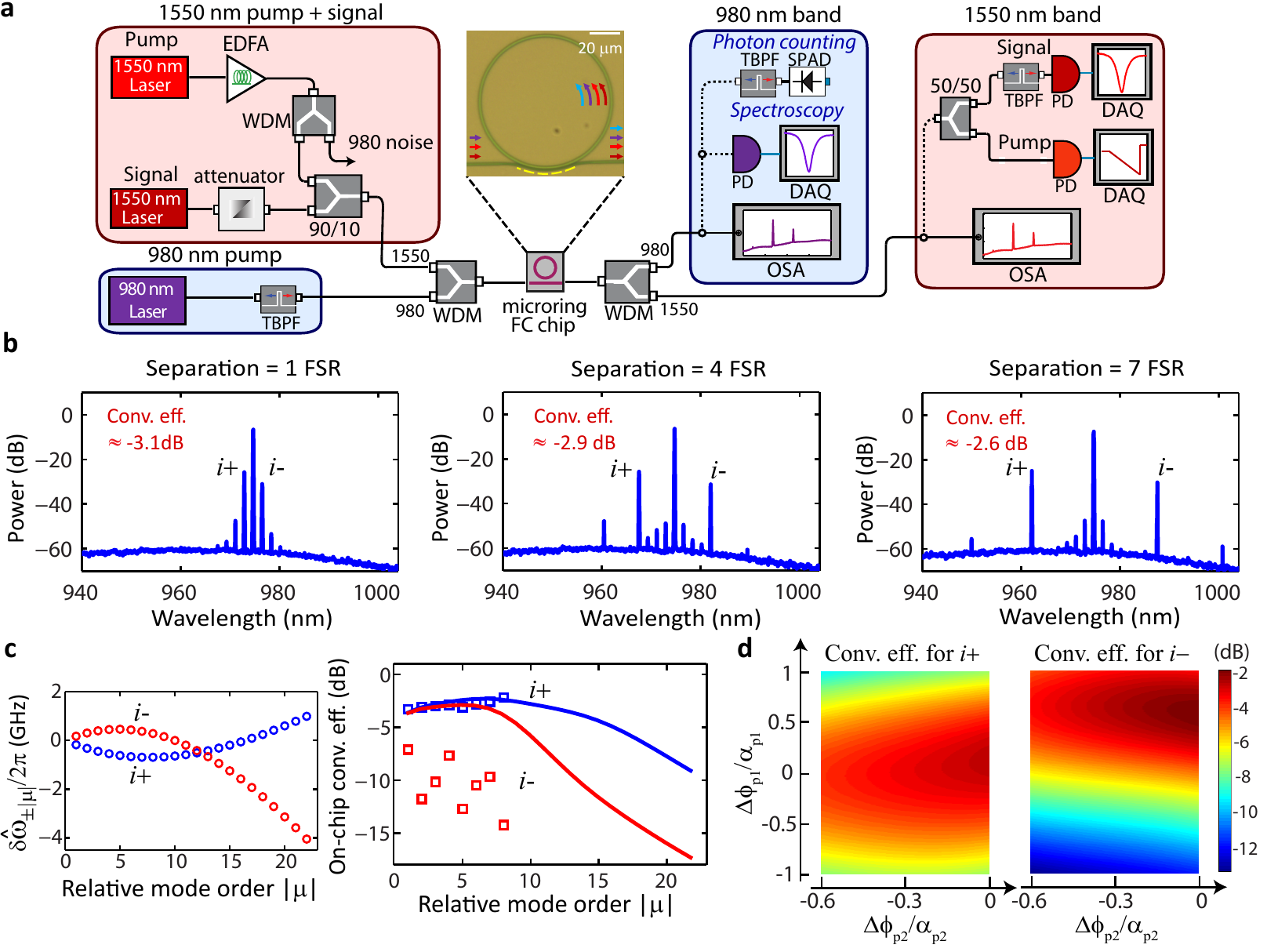}
\caption{\textbf{Experimental setup and additional data for upconversion}. \textbf{a}, Experimental setup for the frequency upconversion from the $1550$ nm band to the $980$ nm band.   \textbf{b}, Representative $980$ nm OSA spectra of upconversion for different frequency separations between the signal and the $1550$ nm pump while keeping the pump configuration fixed: the two pumps in the $980$ nm and $1550$ nm bands are located at $974.4$ nm (power $\approx 8$ mW) and $1564.5$ nm (power $\approx 50$ mW), respectively. A power of 0 dB is referenced to 1 mW.  \textbf{c}, The left figure shows the calculated frequency detunings for $i+$ and $i-$ at different idler positions, and the right figure compares the measured (markers) and simulated (solid lines) on-chip conversion efficiencies for $i+$ and $i-$ as we vary the separation between the signal and the $1550$ nm pump. The error bars for the measured conversion efficiencies due to estimation uncertainties are less than the marker size. In the simulation, $\Delta \phi_{p1}/\alpha_{p1} = 0 $ and $\Delta \phi_{p2}/\alpha_{p2} =-0.1$, where $\alpha_{p1}$ and $\alpha_{p2}$ are the cavity loss parameters for the $980$ nm and $1550$ nm pump, respectively. \textbf{d}, Simulated conversion efficiency for $i+$ (left) and $i-$ (right) as a function of the two pump detunings, $\Delta \phi_{p1}$ ($980$ nm pump) and $\Delta \phi_{p2}$ ($1550$ nm pump).}
\label{Fig_Upconv}
\end{center}
\end{figure*}
\end{center}
\section{$1550$ nm to $980$ nm upconversion: additional discussions}
The experimental setup for the $1550$ nm to $980$ nm upconversion is shown in Fig.~\ref{Fig_Upconv}a, where a $1550$ nm laser is amplified by the EDFA (pump), combined with another $1550$ nm laser (signal), and then combined with the $980$ nm pump before being coupled to the frequency conversion chip. Figure \ref{Fig_Upconv}b shows some representative OSA spectra in the $980$ nm band as we vary the spectral separation between the signal and the $1550$ nm pump, while keeping the pump configuration the same as the example shown in Fig.~5b in the main paper (which corresponds to $8$ FSR separation). By examining Eq.~\ref{Eq_upphia2} (effective detuning of $i+$ in upconversion) and Eq.~\ref{Eq_downphic} (effective detuning of $i+$ in downconversion), we find the frequency detuning of $i+$ is essentially the same for the upconversion and downconversion schemes (except that the sign is opposite), suggesting similar performance for $i+$ between the two cases in terms of conversion efficiency and bandwidth. This is confirmed by comparing the experimental data to the simulation results based on the coupled mode equations as discussed in Section I.B (Fig.~\ref{Fig_Upconv}c). On the other hand, the frequency detuning for $i-$ (Eq.~\ref{Eq_upphib2}) is significantly different from the downconversion case (Eq.~\ref{Eq_downphid}), which seems to indicate that $i-$ should have similar conversion efficiencies as $i+$ even for reasonably large $|\mu|$ values. However, the measured conversion efficiencies are much worse than the simulation results (Fig.~\ref{Fig_Upconv}c). The discrepancy can be attributed to several factors. First, some resonances accommodating $i-$ have been shifted in the spectrum by mode interactions (such as $\mu=-2,-3$, see Fig.~2f in the main paper), which cause large frequency detunings that are not accounted for in the simulation. Second, the transmission at certain resonances (such as $\mu=-8$, see Fig.~\ref{Fig_Trans}c) is much worse than other resonances, possibly due to some resonant scattering from defects along the waveguide or on the cleaved facets, leading to less power in the transmitted idlers. Finally, the frequency detuning of the two pumps can drift during the experiment, which impacts the two idlers in a different way. The $1550$ nm pump is always blue detuned (i.e., $\Delta \phi_{p2} <0$), since otherwise the thermal lock will be lost and no conversion efficiency will be observed. On the other hand, the $980$ nm pump exhibits a Lorentzian transmission scan due to the fact that its power is much smaller compared to the $1550$ nm pump. Though in the experiment we try to set $\Delta \phi_{p1} \approx 0$ by parking the laser to its minimum of transmission, it has been observed that the $980$ nm pump can often drift away from the initial position within a relatively short time. As can be seen from the simulation shown in Fig.~\ref{Fig_Upconv}d,  $i-$ is much more sensitive to the pump detunings than $i+$ for the device under study, especially to the detuning of the $980$ nm pump.

\section{$980$ nm to $1550$ nm downconversion: additional discussions}
\begin{center}
\begin{figure*}[h!]
\begin{center}
\includegraphics[width=0.85\linewidth]{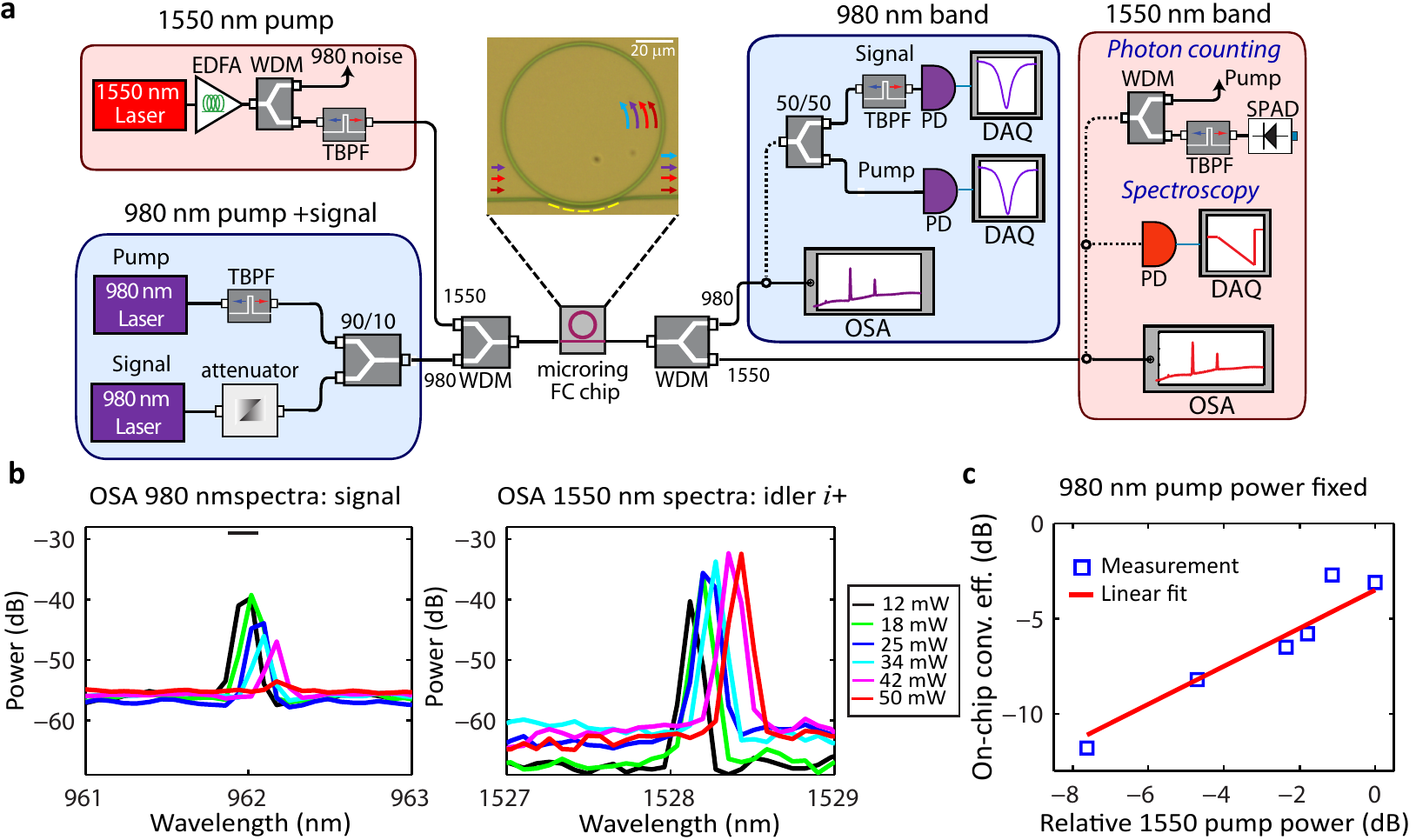}
\caption{\textbf{Experimental setup and additional data for downconversion}. \textbf{a}, Experimental setup for the frequency downconversion from the $980$ nm band to the $1550$ nm band. \textbf{b}, Superimposed OSA spectra for the $980$ nm signal (left) and the $1550$ nm idler $i+$ (right) corresponding to various $1550$ nm pump powers (located at $1559.8$ nm) and a fixed $980$ nm pump power around $8$ mW (located at $974.4$ nm). The small horizontal bar in the left figure marks the signal off-resonance power ($\approx -29$ dB), where a power of $0$ dB is referenced to 1 mW. \textbf{c}, Extracted conversion efficiencies for $i+$ from \textbf{b} (markers) plotted against a linear fit (solid line), showing that the conversion efficiency is proportional to the $1550$ nm pump power. The error bars for the measured conversion efficiencies due to estimation uncertainties are less than the marker size. A power of $0$ dB in the $x$ axis is referenced to the maximum power used in the experiment ($50$ mW).}
\label{Fig_Downconv}
\end{center}
\end{figure*}
\end{center}

The experimental setup for the $980$ nm to $1550$ nm downconversion is shown in Fig.~\ref{Fig_Downconv}a, where we have added a narrowband filter (bandwidth $\approx$ 100 GHz) after the EDFA to remove the ASE noise. In addition, cascaded WDM filters have been used at the detection side to reject the $1550$ nm pump ($>120$ dB suppression) while allowing idlers to pass with a small insertion loss ($\approx 3$ dB). The transmitted idlers are bandpassed by a narrowband filter with a tunable bandwidth ($32$ pm to $600$ pm, , see Fig.~\ref{Fig_Noise}b for its transmission response) and center wavelength ($1460$ nm to $1560$ nm) before detection by a SPAD. Figure \ref{Fig_Downconv} shows the OSA spectra by varying the $1550$ nm pump power while keeping the $980$ pump power fixed. (The pump and signal configuration is the same as the example shown in Fig.~5d in the main paper.) As can be seen, the extinction ratio of the signal resonance increases with the $1550$ nm pump power (left figure), while the generated idler $i+$ increases initially and then saturates (right figure). The summarized conversion efficiency of $i+$ versus the $1550$ nm pump power shown in Fig.~\ref{Fig_Downconv}c indicates the conversion efficiency scales linearly with the $1550$ nm pump power, which is expected as the the conversion efficiency is determined by the product of two pump powers (see Eq.~\ref{Eq_MW})~\cite{ref:Agha_OE_FWM_BS_SM}.

\section{Background noise measurements}
\begin{center}
\begin{figure*}[h!]
\begin{center}
\includegraphics[width=0.85\linewidth]{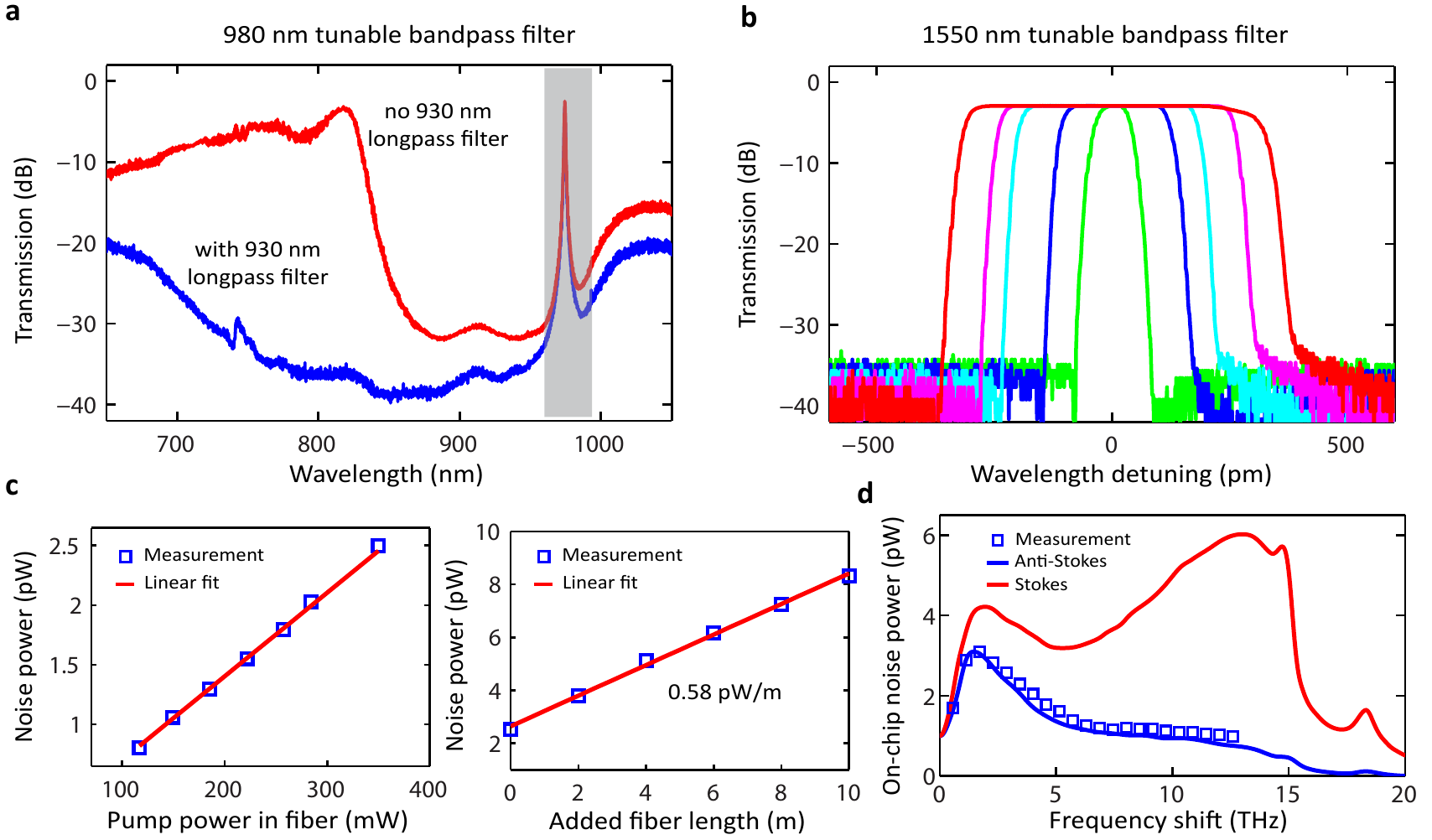}
\caption{\textbf{Filter responses and additional data on background noise measurement}. \textbf{a}, Filter response of the tunable bandpass filter used for the background noise measurement in the $980$ nm intraband conversion experiment (Fig.~4e in the main paper). The passband that is used for the selection of the idler is highlighted. The blue and red curve corresponds to the response of the filter with and without a $930$ nm longpass filter, respectively.  \textbf{b}, Representative filter responses of the bandpass filter used for the background noise measurement in the downconversion experiment (Figs.~6c and 6d in the main paper), showing that its bandwidth can be tuned from $80$ pm to $600$ pm with almost the same insertion loss. \textbf{c}, Background noise measured at the output of the narrowband filter after the EDFA at the input side (see Fig.~\ref{Fig_Downconv}a) as a function of the pump power in the fiber (left) and added fiber length for a fixed pump power around $350$ mW in the fiber (right). On the detection side, the bandpass filter is set at $1528$ nm (pump located at $1559.8$ nm) with a filtering bandwidth of $80$ GHz for both cases. \textbf{d}, The markers are the measured on-chip noise when the $1550$ nm pump is off resonance for a detection bandwidth of $80$ GHz (data re-plotted from Fig.~6c in the main paper as a function of the frequency shift from the pump), while the solid lines are the spontaneous Raman noise spectra from fibers for Stokes (red) and anti-Stokes (blue) sidebands ~\cite{ref:Agrawal_NFO_SM,ref:Lin_photon_pair_OE_SM,ref:Lin_Raman_PRA_SM}.  The error bars in \textbf{c}-\textbf{d} are smaller than the marker size. }
\label{Fig_Noise}
\end{center}
\end{figure*}
\end{center}

In this section, we provide additional data on the background noise measurements performed in this work. Figure \ref{Fig_Noise}a shows the measured transmission response (blue curve) of the tunable bandpass filter used in the noise measurement for the $980$ nm intraband conversion (Fig.~4e in the main paper), which consists of a $980$ nm bandpass filter with a tunable center wavelength from $970$ nm to $990$ nm and a fixed bandwidth of $0.4$ nm, and a longpass filter which suppresses light below $930$ nm ($\approx 1$ dB insertion loss to light with wavelengths $> 930$ nm). It is found that when the $1550$ nm pumps are on resonance, the measured noise by the SPAD can increase significantly (more than tenfold increase) if we remove the $930$ nm longpass filter, indicating the majority of the noise is below the wavelength of $930$ nm. Inspecting the $980$ nm bandpass filter with a supercontinuum white light source and an OSA reveals that the bandpass filter has a wide passband below $820$ nm (red curve in Fig.~\ref{Fig_Noise}a), suggesting the excessive noise (without the $930$ nm longpass) is likely to have wavelengths below $820$ nm. The investigation of its exact nature is the subject of future study.

Figure \ref{Fig_Noise}b shows the transmission responses of the narrowband filter used for the noise measurement in the frequency downconversion experiment (Figs.~6c and 6d in the main paper), which is used to bandpass filter the downconverted idler in the $1550$ nm band before the detection by a SPAD. The filter has a tunable bandwidth of $32$ pm to $600$ pm (although below $80$ pm the insertion loss starts to increase), and a tunable center wavelength from $1460$ nm to $1560$ nm. As mentioned in the main paper, when the $1550$ nm pump is off resonance, we find a broadband noise that scales with the detection bandwidth and is linearly dependent on the pump power. Moreover, the measured noise is almost the same if we remove the chip and introduce a similar insertion loss between the two lensed fibers, strongly suggesting the noise is from the fiber itself. Figure \ref{Fig_Noise}c measures the noise at a fixed frequency shift ($\approx 4$ THz) from the pump at the input side (no chip), showing that the measured noise is linearly proportional to the pump power in the fiber (left figure) and increased fiber length for a fixed pump power (right figure). The extracted ratio of the increased noise power with the fiber length ($\approx 0.58$ pW/m) is consistent with the anti-Stokes Raman signal for a pump power of $350$ mW with a detection bandwidth of $80$ GHz ~\cite{ref:Agrawal_NFO_SM}. To further support this argument, we re-plot the pump-off-resonance noise shown in Fig.~6c in the main paper in Fig.~\ref{Fig_Noise}d (markers) as a function of the frequency shift, and compare it to the spontaneous anti-Stokes Raman spectrum (blue line, multiplied by a factor to match its peak power to that of the pump-off-resonance noise). The excellent agreement of the spectral shape confirms that the pump-off-resonance noise is mostly likely the anti-Stokes Raman signal generated inside the fiber.

\end{document}